\documentstyle[aps,epsf]{revtex}
\def\gtrsim
{\raise2pt\hbox
{$\displaystyle{}\mathrel{\mathop>_{\raise1pt\hbox{$\sim$}}}{}$}} 
\def\lesssim
{\raise2pt\hbox
{$\displaystyle{}\mathrel{\mathop<_{\raise1pt\hbox{$\sim$}}}{}$}}
\begin{document}
\title{Effects of Domain Wall on 
Electronic Transport Properties in Mesoscopic Wire of
Metallic Ferromagnets 
} 
\author{Gen { Tatara}\\
Max Planck Institut fur Mikrostrukturphysik, Weinberg 2, D-06120 Halle,
Germany\\ and \\
Graduate School of Science, Osaka University, Toyonaka, Osaka 560-0043}
\sloppy
\maketitle
\begin{abstract}
We study the effect of the domain wall on electronic transport 
properties in wire of ferromagnetic 3$d$ transition metals based on  
the linear response theory. 
We considered the exchange interaction between the conduction electron 
and the magnetization, taking into 
account the scattering by impurities as well. 
The effective electron-wall interaction is derived by use of a local 
gauge transformation in the spin space.
This interaction is treated perturbatively to the second order.
The conductivity contribution within the classical 
(Boltzmann) transport theory turns out to be negligiblly small in bulk 
magnets, due to a large thickness of the wall compared with the fermi 
wavelength.  
It can be, however, significant in ballistic nanocontacts, as indicated in 
recent experiments.
We also discuss the quantum correction in disordered case where the quantum 
coherence among electrons becomes important. In such case of weak localization
the wall can contribute to a decrease of resistivity by
causing dephasing.
At lower temperature this effect grows and 
can win over the classical contribution, in 
particular in wire of diameter $L_{\perp}\lesssim \ell_{\varphi}$, 
$\ell_{\varphi}$ being the inelastic diffusion length. 
Conductance change of the quantum origin 
caused by the motion of the wall is also discussed. 
\end{abstract}
\newcommand{\kv}{{\bf k}}
\newcommand{\qv}{{\bf q}}
\newcommand{\pv}{{\bf p}}
\newcommand{\xv}{{\bf x}}
\newcommand{\Xv}{{\bf X}}
\newcommand{\np}{n'}
\newcommand{\kvm}{{\bf k}-\frac{q}{2}}
\newcommand{\kvp}{{\bf k}+\frac{q}{2}}
\newcommand{\omegap}{\omega'}
\newcommand{\tilDelta}{\tilde{\Delta}}
\newcommand{\tilGam}{\tilde{\Gamma}_{0}}
\newcommand{\till}{\tilde{l_{\sigma}}}
\section{Introduction}
\subsection{Resistivity in bulk magnetic metals}

Since more than a century ago number of studies has been carried out 
on 
the electric transport properties in ferromagnetic metals. 
They revealed many remarkable 
features which are not seen in non-magnetic metals.
One of the most notable phenomena would be the hysteretic and 
anisotropic behavior of 
the resistance in the magnetic field (magnetoresistance) observed at 
small magnetic field of  $\lesssim 1$T, which has been already 
noted more than a hundred years ago\cite{Thomson1857}.
The magnetoresistance in the case of the field $H$ parallel to the 
current $I$
takes a minimum at a finite value of the field ($\sim200$Oe for 
instance for 
the case of Ni and Fe) .
If the field is applied perpendicular to the current, the curve of 
magnetoresistance is reversed; 
namely resistivity shows a maximum at certain field and 
decreases as the field deviates from the value.
The hysteretic behavior of the magnetoresistance is due to the fact 
that the resisitivity is mostly governed by 
the magnetization $M$, which corresponds to a field of $\sim 1$T and
thus is larger than the applied magnetic field in 
the field range we are interested in, except very close to the 
coercive field where the magnetization vanishes. 
The observed resistivity $\rho$ as a function of the field, $H$, has 
been shown to be well fitted by a phenomenological relation of
$\rho \sim \rho_{0}+\Delta\rho_{\rm ani} <\cos^{2} \theta_{M}>$,
where $\rho_{0}$ is the field independent part 
and $\Delta\rho_{\rm ani}$ measures the strength of the 
anisotropy in the resistivity\cite{Smit51,McGuire75} .
$\theta_{M}$ is the mutual angle between the local magnetization and 
the current, which depends on the magnetic field, 
and bracket denotes the average over the sample.
The anisotropy $\Delta\rho_{\rm ani}$ is expressed in terms of 
the resistivity in the case of the field parallel to the 
current, $\rho_{\parallel}$, and that in the perpendicular case, 
$\rho_{\perp}$, as 
$\Delta\rho_{\rm ani}=\rho_{\parallel}-\rho_{\perp}$.
In most of the ferromagnetic metals $\rho_{\parallel}$ takes a larger 
value than $\rho_{\perp}$; namely $\Delta\rho_{\rm ani}$ is 
positive\cite{RefKent}.
This anisotropic behavior of the resistivity is called anisotropic 
magnetoresistance (AMR).

A microscopic explanation of AMR has been given by 
Smit\cite{Smit51} and McGuire and Potter\cite{McGuire75} in the following 
way.
They discussed that the conduction ($s$-) electron is coupled to the 
magnetization due to the scattered into the
magnetic ($d$-) band and spin-orbit interaction there, and that this 
process gives rise to an spin asymmetric lifetime, which depends on the 
angle between the current and the field, resulting in an anisotropy.
According to their arguments, positiveness of $\Delta\rho_{\rm ani}$ is 
explained if the resistivity is dominated by the electron of the 
minority spin.
It was shown recently that the magnitude and the sign of $\Delta\rho_{\rm ani}$ 
can indeed be controlled by changing the 
spin asymmetry in scattering in magnetic multilayers, which would support 
the above explanation\cite{Hsu97}. 

\subsection{Domain wall contribution to classical magnetoresistance}

The magnetoresistance observed so far in bulk 
materials is mostly understood well in terms of AMR effect, which 
assumes that the magnetization changes very slowly and hence the 
electron feels only the average magnetization\cite{McGuire75}.
However this assumption may not be good in real magnets which 
contains many domains having different direction of the local 
magnetization. 
In fact the boundary of these domains is a structure called a domain wall 
where the spins 
rotate spatially within a finite distance (Fig. \ref{FIGDW}).
Such domain walls
can lead to a scattering of the electron, which 
is not taken into account in the standard AMR argument.
In this paper we exclusively consider the effect of domain walls on the 
electronic transport properties.
Of particular interest would be the case of a small sample less than a 
typical domain size $\simeq 1\sim10\mu$m, since there the magnetic 
properties can be described in terms of domain wall configuration.
In fact in such samples the magnetization process as the field is swept 
will be described by the 
nucleation of one or a few domain walls, motion of the walls by 
depinning followed by the annihilation. 
If domain walls can have some contribution to the resistivity these 
dynamics of the wall will appear in the magnetoresistance as discrete jumps 
in the measurement with the field is swept, 
since such events are faster (e.g., the speed of domain wall motion is 
estimated to be about 182m/s in submicron wire of NiFe\cite{Ono99}) 
compared to the sweep speed. 
These effects are
similar to the  Barkhausen noise\cite{Barkhausen19}.
Such effect will not be seen very clearly in bulk samples, since there the 
contribution from many domain walls will be summed up in the 
observed magnetoresistance and thus the contribution from each domain 
wall will not be visible.
Indeed such jumps in the magnetoresistance has been observed  
recently in a Ni wire of about the diameter of 
300\AA\cite{Giordano94,Hong95,Hong96,Hong98},
in submicron\cite{Otani97} and micron size wire of 
Fe\cite{Kent98,Ruediger98,Ruediger98b} and Co 
wires\cite{Ruediger98c,Otani98}.

Let us here give a rough estimation of the effect of the wall on the 
classical electron transport. 
The most important parameter in doing this is the 
thickness of the wall, $\lambda$, the 
length in which the spins rotate spatially between the two adjacent 
domains.
This quantity is determined by the competition between the 
exchange energy per site, $J$, which aligns the neighboring spins and 
the 
magnetic anisotropy energy in the easy axis, $K$, which tends to make 
thinner the wall to keep minimum the deviation of spins from the easy 
axis,
as $\lambda =\sqrt{J/K}a$, $a$ being the lattice constant.
Thus $\lambda$ depends on the material and also on the sample shape 
since $K$ depends on the shape.
In the case of 3$d$ transition metals such as iron and nickel 
$\lambda\simeq 500\sim1000$\AA\cite{Hong95}, and $\lambda\sim 150$\AA\ 
in Co thin film\cite{Gregg96}.
Hence this is very large compared 
with the length scale of the electron, 
$k_{F}^{-1}\sim O(1$\AA$)$ ($k_{F}$ being the Fermi wave length of the 
electron).
In such materials, therefore, the conduction electron can 
adiabatically adjust 
itself to the local magnetization at every point as it passes through 
the wall, resulting in a very small scattering probability by the 
wall. 
The classical resistivity in the Boltzmann's sense, which is proportional to 
the reflection probability by the wall, is thus expected to be negligiblly 
small. 
This was explicitly shown by Cabrera and Falicov\cite{Cabrera74} by 
calculating in the clean limit 
the reflection coefficient based on the one-dimensional 
Schr\"odinger equation for the electron coupled via exchange coupling 
to the magnetization whose configuration is a domain wall. 
They obtained for the case of thick wall, $k_{F}\lambda\gg 1$ the 
expression 
$\rho_{\rm w}\propto \exp(-2\pi k_{F}\lambda)$ for the wall 
contribution, and discussed that the wall can have large  effect only 
if the wall is extremely thin ($k_{F}\lambda \lesssim 1$) and if the 
conduction electron is strongly polarised by the magnetization.
In hard magnets with strong magnetic anisotropy as in materials 
containing rare-earth like SmCo$_{5}$ and manganese perovskites,  
$\lambda$ can be as small as about $10$\AA.
Thus in such systems the resistivity may be dominated by scattering 
by domain walls, as indicated in recent experiments
in the ferromagnetic phase of manganese perovskites\cite{Schiffer95}.

In the past few years there has been a renewal of interest in theory 
of the classical resistivity due to the domain 
wall\cite{Yamanaka96,TF97,Zhang97,Brataas98,Brataas99,vanHoof99,TG99}, because 
precise measurements of the effects are now becoming possible.
It has been pointed out by Levy and Zhang\cite{Zhang97}
that in the presence of 
a spin asymmetry in the electron lifetime, which is the case in most 
ferromagnets, domain wall can have a substantial effect on the 
classical resistivity by mixing the two spin channels with different 
resistivity, in agreement with the experiment on Co film\cite{Gregg96}.
The effect of lifetime asymmetry has been further discussed in detail 
in ref. \cite{Brataas99}. 
The domain wall resistivity in the ballistic limit has been discussed 
by use of realistic band structures in ref. \cite{vanHoof99}.
It was shown there that the existence of nearly degenerate bands at 
the Fermi level in real magnets enhances the classical resistivity 
due to the wall.

One of systems of recent particluar interest is an atomic scale contact 
of magnetic metals in the ballistic region.  
In such narrow contacts the profile of the wall is 
determined mostly by the shape rather than the anisotropy energy of the 
bulk magnets, and thus the wall is trapped in a 
contact region, which is typically nm scale. 
The adiabaticity does not hold in such cases of small $k_{F}\lambda$ 
comparable to 1, and a large effect from the wall is 
expected\cite{Cabrera74,vanHoof99,TG99}.
In fact a magnetoresistance of 200\% has been observed in ballistic Ni 
nanocontacts where the number of the channel is less than 
$N\lesssim10$\cite{Garcia99}, 
and this has been interpreted in terms 
of a strong reflection by a nanometer scale domain wall\cite{TG99}.
Ballistic nano-contacts would be one of the novel systems in which a large 
magneto-electronic effect is expected.
In the dirty case in contrast, the effect is not so large, since 
the wall scattering is smeared out by impurity scattering\cite{TG99}.

Intensive studies has been carried out experimentaly on sub-micron 
scale wires
\cite{Giordano94,Hong95,Hong96,Hong98,Otani97,Ruediger98,Ruediger98b,
Ruediger98c,Otani98,Ono98}. These are
explained in detail in \S\ref{SECwires}.
A structure of layers of hard and soft ferromagnets would be an 
interesting possibility to investigate the effect of the domain wall, 
since there the thickness of the wall can be controlled by the external 
magnetic field as carried out in ref. \cite{Mibu98}. 
Measurement of conductance through a sub-micron ferromagnetic dot on 
a semi-conductor wire has been carried out\cite{Yamada98}.
Semi-conductors are interesting since the electron can feel a larger 
coupling to the wall can be larger due to a larger electron 
Fermi wavelength compared with metals.

\subsection{Quantum transport}
So far these studies both experimentally and theoretically
are carried out mostly in the case of low resistivity materials, 
where the transport can be discussed in terms of the classical theory. 
However besides the classical transport, there 
is another important aspect in electronic transport at low 
temperature. This is the effect of the quantum coherence among  
electrons, which modifies the low energy electronic properties significantly.
The effect becomes important in disordered metals with high 
resistivity, where the elastic mean free path becomes shorter due to 
the elastic impurities, which cause only elastic scattering. 
The electron wave scattered by such normal impurities can interfere 
with the incoming wave, leading to a state like a standing wave. 
This is called weak localization and the resistivity in this case is 
enhanced due to the quantum interference\cite{Bergmann84,Lee85}.
This correction becomes large in low dimensions such as in wires since there 
the interference becomes stronger.
The interesting point of this situation is 
that the electronic properties are very sensitive to a small 
disturbance because of the presence of coherence.
For instance in a non-magnetic metals of micron size, 
even a motion of a {\it single} impurity atom has been shown to
change at low temperature the conductance ($G\equiv \sigma 
A_{\perp}/L$, $A_{\perp}$ and $L$ being the crossssectional area and 
the length of the system) of the entire 
system by disturbing the coherence\cite{Meisen89}. 
The magnitude of the conductance change turns out in most cases 
to be a universal order of $e^{2}/h$\cite{Feng86}.
It is then reasonable to expect that in mesoscopic ferromagnetic 
metals, domain walls can lead to a large change of 
conductance in the presence of quantum interference.
In fact this possibility has been pointed out in ref. \cite{TF97}.
The most remarkable point is that because a domain wall destroys the 
interference among the electron, the wall contributes to a negative 
quantum correction to the resistivity. 
In disordered 3$d$ transition metals, this quantum correction
can win over the classical contribution, and thus a nucleation of a wall 
in the sample may lead to a {\it decrease} of resistivity.
A recent numerical simulation also supports the negative resistivity 
contribution from 
the wall in disordered thin wire\cite{Jonkers99}. 
Recently it has been pointed out that the geometric phase attached to 
the electron spin as it passes through the wall can also cause 
important dephasing effect, which would become important in multiply
 connected geometry\cite{Geller98,Loss99}. 
The effect of dephasing due to the magnetic origin has been 
considered also in the thin film of metal sandwiched by ferromagnetic 
layers. There the internal magnetic field at the interface  causes 
dephasing in the conduction layer in the 
presence of the spin-orbit scattering, which 
contributes to a positive magnetoresistance\cite{TF99}. 

In real materials there are source of dephasing other than such 
magnetic objects, such as phonons, electron-electron 
interaction\cite{Bergmann84,Lee85}. In magnets the fluctuation of the 
magnetization, i.e., spin wave, also contributes to the dephasing.
Such effects can be avoided by observing at low termperature.
In bulk or film of ferromagnetic metals, the existence of strong internal 
field of about 1T can also destroy the coherence\cite{Bergmann84,Lee85}.
Fortunately this is not the case in a narrow wire as discussed in 
\S\ref{SECmagneticfield}. 

Quite recently experimental efforts in search for the quantum 
coherence in magnetic metals have been carried 
out\cite{Giroud98}.
The resistivity of a Co wire of 1000\AA\ width which contains a loop of 
500nm has been studied in ref. \cite{Giroud98}. It was argued there 
that from the absence of Aharaonov-Bohm oscillations the 
phase-breaking length of the Co at $T=0.29$K is shorter than $0.3\mu$m.
The study of a ferromagnetic single electron transistor of double 
Ni/NiO/Co tunnel junctions at 20mK has revealed that higher order tunneling 
processes dominate in the Coulomb blockade region, in which two or more 
electrons coherently tunnels through the two junctions\cite{OnoK97}. This 
result suggests the electronic coherence is kept there for $\sim 2.5\mu$m, 
which is the length of the Co island.

Because of its sensitivity these conductance fluctuation as a consequence of 
quantum interference has already been used
as a probe in the studies of various mesoscopic metallic or semi-conducting 
systems. 
For example a telegraph noise due to a two-level oscillation of a 
defect in Bi film has been investigated and it turned out that the 
oscillation at $T\lesssim 1$K is governed by the quantum tunneling 
subject to the dissipation from the conduction electron\cite{Golding92}. 
Such measurement of the quantum transport properties has been proved to 
be a useful probe also for studies of mesoscopic spin-glass 
systems\cite{Meyer95,Strunk98} and the magnetization flip of 
mesoscopic magnets\cite{Coppinger94}. 

\subsection{Mesoscopic magnets}
Quite recently the resistivity measurements has been carried out to 
investigate the the reversal of the magnetization in mesoscopic 
magnets at low 
temperature\cite{Giordano94,Hong95,Hong96,Hong98,Otani97,
Ruediger98,Ruediger98b,Ruediger98c,Otani98,Ono98,Mibu98,Yamada98}. 
They are motivated by the interest in the magnetization reversal via 
quantum fluctuation of magnetization\cite{Chichilianne94}. 
In the case of very small magnets 
this process of magnetization flip is expected to be well described by 
the uniform rotation of the total magnetization.
At high temperature and in bulk magnets 
these processes are caused by the thermal activation over the energy 
barriers due to the anisotropy energy, nucleation energy or the 
pinning potential. 
If the thermal activation is the only process of magnetization 
reversal, a freezing of the magnetization is expected in the 
limit of zero temperature.
However, particles of nanometer size have a finite quantum fluctuation, 
although it is not very large in large magnets compared 
to the atomic scale.
Thus  another process of a magnetization flip is possible, {\it i.e.}, 
via a quantum tunneling, leading to
a finite relaxation rate even at zero 
temperature\cite{Chudnovsky88,Chichilianne94}.
The tunneling entity, the total magnetization 
in nanoscale magnets,  are macroscopic or semimacroscopic 
objects compared to the atomic scale, and thus 
these tunnelings  are  
called a macroscopic quantum tunneling (MQT)\cite{CL81}.
Assuming the case of very small magnets, the theories so far are based on 
a simplified picture of the relaxation being described by a single 
variable which tunnels through a single energy barrier.
Then the relaxation expected will depend exponentially on time, 
$M\propto \exp(-\Gamma t)$ ($\Gamma$ being the decay rate).
Many experimental attempts has been made in search for such quantum 
tunneling of magnetization for about a decade\cite{Chichilianne94}.
In the studies at the beginning of the 90's the the relaxation of the 
magnetization of small ferromagnetic particles whose diameters are 
about $40\sim150$\AA\ has been measured by use of SQUID 
(superconducting quantum interference device)\cite{Balcells92,Barbara93}.
Because the sensitivity of the SQUID used was not so high, only a 
collection of many particles could be measured. 
The observed relaxation was logarithmic in a wide range of the time, 
$M\sim S_{\rm v}\ln t$, where $S_{\rm v}$ is a constant called the 
magnetic viscosity, in contrast to the exponential dependence 
predicted by simple theories.
This discrepancy would be due to the distribution of the size and 
shape of the particles. 
A tendency of saturation of $S_{\rm v}$ to a finite value as 
temperature 
decreases has been reported, but 
because the size distribution is unknown, 
no definite conclusion could be drawn concerning whether the 
relaxation is really of the quantum origin or not.

Recent experiments thus aim at observation of relaxation without 
any ambiguity in the interpretation.
A first such possibility would be to use better samples. 
For example, magnetic molecules are good candidates because of their
uniform structure. 
In fact a crystal of Mn$_{12}$ acetate, made up of 
molecules containing twelve manganese atoms which form the total spin 
of $S=10$, has been shown to exhibit the exponential decay of 
magnetization\cite{Paulsen95} consistent with the picture of a single 
energy barrier\cite{Chudnovsky88}.  
The data clearly showed that the decay is due to the quantum 
tunneling below $2$K.
Further study on the system revealed the resonant tunneling in each 
molecule\cite{Friedman96}.
These behavior consistent with a picture of a tunneling of the total 
magnetization through a single energy barrier may be natural because 
the spin which tunnels, $S=10$, is microscopic rather than 
mesoscopic. 
Magnets of biological origin is another candidate for definitive 
experiments.
For example the AC susceptibility has been measured on ferritine 
particles 
from horse spleen, each particle is spherical and contains a cluster 
of antiferromagnetically coupled Fe atoms, the diameter being 
$75$\AA\cite{Awaschalom93}.
The imaginary part of the susceptibility showed a sharp peak at about 
1MHz, which may to be explained by the resonance with the coherent 
oscillation of the  magnetization due to the quantum tunneling, 
namely a macroscopic quantum coherence. 
Although there has been some arguments concerning the 
interpretation\cite{Garg93}, further studies on samples with different 
diameter indicates the above explanation would be  
correct\cite{Gider95,Tejada97}.

As a second approach to a definitive experiments, direct 
observation of the magnetization of a single nanoscale magnet has 
recently been carried out.
Magnetic force microscope (MFM)\cite{Schultz94}  and 
micro-SQUID\cite{Wernsdorfer95} has been used to study
the reversal of the magnetization of 
a small particle of $\gamma$-Fe$_{2}$O$_{3}$ with the 
size of $3000$\AA\ at room temperature and of $800$\AA\ Co particle
at low temperature.
It turned out that even for such a small particle the relaxation is 
logarithmic in time, which is argued to be due to pinning by inherent 
defects 
and surfaces\cite{Schultz94,Awaschalom95}.
Indeed recent measurement on a smaller particle of $150\sim 
300$\AA\ indicates the uniform rotation of the 
magnetization\cite{Wernsdorfer97}. In the observed temperature range  
of $0.2<T<6$K, however, there was no indication of quantum tunneling.

In the case of wires, the nucleation, depinning and annihilation of 
domain walls can be driven by the quantum 
fluctuation\cite{Stamp91,TF94}.
Intensive efforts has been put for experimental confirmation on Ni wires 
of width of $300\sim400$\AA\ by use of resistivity 
measurements\cite{Giordano94,Hong95,Hong96}. 
These earlier works seemed to suggest the depinning by quantum tunneling 
below 5K\cite{Giordano94,Hong95,Hong96}, but in the SQUID measurement 
of a 650\AA\ wire no evidence of quantum tunneling has been found in 
the temperature range of $0.1\sim6$K\cite{Wernsdorfer96}.
It has been, however,  indicated in the observation of the response to the 
microwave that the energy levels of a pinned wall is quantized\cite{Hong96}.
Another possibility of tunneling of the chirality of the domain wall, 
i.e., the way the magnetization rotates inside the wall, has been 
discussed in Refs. \cite{Braun96,TT96}. 
\subsection{Resistivity measurements of magnetic wires}
\label{SECwires}
For studies of the magnetic behaviors of nanoscale metallic 
magnets, electronic transport measurement can be effective 
because of its sensitivity and facility.
In fact a telegraph noise measurement of an antiferromagnetically coupled
small cluster of ErAs on GaAs, where the diameter of the clusters are 
estimated to be about $30$\AA\ has been carried out\cite{Coppinger94}.
It was shown that the telegraph noise was due to a flip of a 
magnetization of each cluster, the magnitude being 
$\sim 41\sim 150\mu_{B}$. Result indicated that the flip is due to the 
tunneling below $350$mK.
Transport measurements are particularly useful in magnetic wires.
The measurement of the magnetoresistance of a Ni wire with diameter 
of $200\sim400$\AA with the length of $\sim 10\mu$m has revealed a 
discrete jump of the resistivity as the magnetic field along the wire 
direction is swept at temperature of $1.4\sim22$K\cite{Giordano94,Hong95}.
The global curve of magnetoresistance observed there
was similar to the behavior of
positive magnetoresistance common in bulk ferromagnetic 
metals\cite{McGuire75}.
The jump appeared close to the minimum of the resistivity
and was reproducible except for a slight distribution of the value of 
the field at which the jump occurs.
The magnitude was about 0.02\% of the total resistivity.
It was argued there that the jump would be due to the change of the 
average magnetization associated with the depinning of a domain wall.
The explanation of this jump by use of the conventional AMR\cite{McGuire75}
has been given there and 
the displacement of the wall was estimated from the magnitude of the 
jump to be about $1.2\mu$m.
They argued from the distribution of the position of the jump 
that the depinning is dominated by quantum 
tunneling below $5$K\cite{Hong95}, but another measurement on 650\AA\ 
carried out by use of SQUID has found only process of thermal activation 
down to 0.13K\cite{Wernsdorfer96}. 
Recently similar experiment was carried out on Fe wire with width of 
3000\AA\ and there two reproducible 
jumps in the resistivity have been observed at 
0.3K, negative jump followed by a positive one as the field 
increases\cite{Otani97}. The first and second jump are expected to 
be due to the 
nucleation and annihilation of a domain wall, respectively, and then 
the resistivity seems to be suppressed by the wall. 
This decrease of resistivity by the wall is expected when the 
coherence among electrons is important\cite{TF97}.
Other recent measurements on sub-micron scale wires also suggest a 
negative contribution of the wall at low 
temperature\cite{Hong98,Otani98}.
These samples are dirty with the residual resistivity of about 
$\rho_{0}\sim 10\mu\Omega$cm. Negative contribution has also been seen in 
cleaner sample of Fe with $\rho_{0}\sim0.2\mu\Omega$cm\cite{Ruediger98}.
These effects have been observed to grow at lower temperature 
(below 50K\cite{Ruediger98} and 20K\cite{Otani98}).
These behaviors might be considered as an
experimental confirmation of the quantum coherence effect predicted 
in ref. \cite{TF97}, 
although further detailed studies are obviously needed.
For instance the temperature of 20K seems to be rather high for the 
weak localization correction to be important, especially for the 
clean case of ref. \cite{Ruediger98}.
A classical origin of negative contribution from the wall, arising 
from the effect of the surface, has 
been suggested in ref. \cite{Ruediger98b}.
Thus at present the reduction of the resistivity due to the wall 
observed in wires at low temperature is not fully understood.
In contrast, at room temperature, the existence of 
the walls has been shown to enhance the resistivity 
in wires\cite{Ono98} and films\cite{Gregg96}, which is 
interpreted as classical process of reflection with the lifetime 
asymmetry taken into account\cite{Zhang97}. 

In order to give an definitive interpretation for such discrete jumps in the 
magnetoresistance, calculation of the resistivity in the presence of 
a domain wall based on a microscopic consideration which takes into 
account various effect such as AMR, the quantum correction and the spin 
asymmetry is necessary.
In this paper we focus on the effect of the 
quantum correction, since it can be especially 
interesting because of a substantial enhancement of the effect of the wall 
expected in low dimension and at low temperature. 
A resistivity measurement in the weakly localized regime of 
the electron is expected to be a powerful tool for studies of 
nanoscale magnets at low temperature.

This paper is organized as follows.
In section \ref{SECclassical} the classical resistivity due to a 
domain wall in the Boltzmann's sense is calculated based on the 
theory 
of linear response taking account the impurity scattering at the same time.
The electron is treated as in three-dimensions since in actual wires 
the diameter $L_{\perp}$ would be large enough, 
$L_{\perp} \gg k_{F}^{-1}$.
The wall is treated as flat which is justified if $L_{\perp}\lesssim 
\lambda$. 
The interaction between the conduction electron and the 
magnetization is the exchange coupling.
We rewrite the system by use of a local gauge transformation which 
makes the $z$-axis of the electron spin chosen always along the local 
magnetization.
The wall is then expressed by a classical gauge field, if the 
reaction 
from the electron on the wall is neglected. 
The effect of the wall is taken into account up to the second order.
In section \ref{SECquantum} the quantum correction to the resistivity 
is discussed.
The quantum correction becomes important when there is enough 
scattering by disorder, which causes the interference between the 
scattered and unscattered electron waves, leading to weakly 
localized states of the electron.
This interference effect is expressed by a particle-particle ladder 
(Cooperon), which is singular at small momentum transfer.
The domain wall in ferromagnetic metals turns out to give rise to a 
mass in the Cooperon, thus it suppresses the localization.
The resistivity will, therefore, decrease when a domain wall is 
nucleated, in contrast to naive classical intuition.
The effect would be particularly important in narrow wires.
The resistivity change due to the motion of the wall is discussed in 
section \ref{SECCF}.
The case of weakly localized electron system is considered since 
there 
the effect will be much larger than that expected from the classical 
argument based of the anisotropic magnetoresistance.
Summary is given in section \ref{SECsummary}.
Details of calculation is given in Appendices.
This paper is an extended and detailed paper of the letter published 
before\cite{TF97}.

\section{Classical Resistivity due to a domain wall}
\label{SECclassical}
We consider here the case of the $s$-$d$ model where the conduction 
electron, $c$, couples to a localized spin, $\mbox{\boldmath{$S$}}$, 
by exchange coupling, the coupling constant being $g$. 
The Lagrangian of the electron in the imaginary time (denoted by 
$\tau$) is then written as
\begin{equation}
L=\sum_{{\bf k}{\sigma}}
c^{\dagger}_{{\bf k}{\sigma}}
(\partial_{\tau}+\epsilon_{{\bf k}})c_{{\bf k}{\sigma}}
-g\int d^{3}x\mbox{\boldmath{$S$}}({\bf x})(c^{\dagger}
\mbox{\boldmath{$\sigma$}}c)+H_{\rm imp}, \label{L}
\end{equation}
where 
$\epsilon_{{\bf k}}\equiv 
\hbar^{2}{\bf k}^{2}/2m-\epsilon_{F}$
($\epsilon_{F}$ being the Fermi energy). 
The spin index is denoted by 
$\sigma=\pm$ and $\mbox{\boldmath{$\sigma$}}$ is the Pauli matrix.
The last term describes the interaction between the electron and the 
impurity:
\begin{equation}
H_{\rm imp}\equiv v\sum_{\kv,\qv} \sum_{i=1}^{n_{\rm i}} e^{i\qv \cdot \Xv_{i}} 
	c_{\kv+\qv}^{\dagger} c_{\kv},
	\label{Vimp}
\end{equation}
where $v$ is the coupling constant and $\Xv_{i}$'s are the position of the impurities
which is assumed to be random, $n_{\rm i}$ being the number of  impurities.
Due to this interaction the electron has a finite elastic lifetime, 
$\tau\equiv \hbar/[2\pi n_{\rm i}v^{2}N(0)]$, 
$N(0)\equiv Vmk_{F}/2\pi^{2}\hbar^{2}$ being density of states at 
Fermi energy ($V$ is the system volume).
Here we assume the same lifetime for electron with both majority 
and minority spin.
The effect of the lifetime difference has been discussed in 
Refs.\cite{Zhang97,Brataas98,Brataas99}.

In the case of the system described by a single-band model, it can be 
shown in the Hartree-Fock approximation 
that the effective Lagrangian describing the low energy behavior of 
the electron and the magnetization 
is the same as eq. (\ref{L}) with $g$ replaced by the 
Coulomb repulsion, $U$, but then the 
exchange coupling $J$ below is determined by $U$\cite{TF94}.
The Lagrangian (\ref{L}) can describe both $s$ and $d$ electron if $g$ 
and thus the splitting ($\Delta$) below is appropriately chosen.

The configuration of the localized spin is determined by the 
ferromagnetic Heisenberg model,
\begin{equation}
H_{S}=\int d^{3} x \left[ 
\frac{J}{2}|\nabla \mbox{\boldmath{$S$}}|^{2}
-\frac{K}{2}S_{z}^{2} +\frac{K_{\perp}}{2}S_{x}^{2} \right], \label{HM}
\end{equation}
where $J$ is the effective exchange energy
and $K$ and $K_{\perp}$ are the magnetic anisotropy energy introduced 
phenomenologically, which are assumed to include the effect of
the demagnetization field.
The transverse anisotropy energy, $K_{\perp}$, which determines the 
inertial mass of the wall, does not affect our 
calculation of a static wall. It becomes essential in discussing the 
dynamics of the magnetization such as the tunneling of the 
chirality\cite{Braun96,TT96}.

Here we are interested in the solution of a single domain wall. 
In terms of the polar coordinates, $(\theta,\phi)$, that represents 
the 
direction of $\mbox{\boldmath{$S$}}$, the solution of a domain wall 
at $z=z_{0}$ is given by
\begin{equation}
	\cos\theta=\tanh\frac{z-z_{0}}{\lambda}, \;\; \phi=\pm\frac{\pi}{2}.
	\label{DWsolution}
\end{equation}
Here $\lambda=\sqrt{K/J}$ is the thickness of the wall.
The configuration is depicted in Fig. \ref{FIGDW}.

In eq. (\ref{L}) the last term represents the interaction between the 
local spin and the electron. 
This term, however, contains the interaction, which exists even
far from the wall, and so does not lead to 
the scattering of the electron. 
Thus we need to rewrite this interaction to the form appropriate for 
the 
perturbative calculation of resistivity.
This can be carried out by use of the local gauge transformation 
which rotates the spin axis of the electron so that the $z$-axis is 
always along the local direction of the localized spin $\mbox{\boldmath{$S$}}$.
Explicitly this transformation is given by 
\begin{equation}
	a_{\sigma}=\sigma \cos\frac{\theta}{2}\; c_{\sigma}
	+e^{-i\sigma\phi}\sin\frac{\theta}{2}\; c_{-\sigma}.
	\label{gaugetr}
\end{equation}
In terms of the new electron operator, $a$, the Lagrangian is written 
as\cite{TF94}
\begin{equation}
L=\sum_{{\bf k}{\sigma}}
a^{\dagger}_{{\bf k}{\sigma}}
(\partial_{\tau}+\epsilon_{{\bf k}\sigma})a_{{\bf k}{\sigma}}
+H_{\rm int}+H_{\rm imp},
	\label{Lnew}
\end{equation}
where 
\begin{equation}
	\epsilon_{{\bf k}\sigma}
\equiv\epsilon_{{\bf k}}-\sigma \Delta
	\label{energynew}
\end{equation}
is the energy of the electron polarised in $z$-direction by the 
magnetization
with $\Delta\equiv g |\mbox{\boldmath{$S$}}|$ being half the 
splitting 
between the up and down spin electrons. 
The term $H_{\rm int}$ describes the interaction between the electron 
and the wall and is given by
\begin{eqnarray}
H_{\rm int} &\equiv&  \frac{\hbar^{2}}{2m} \int d^{3}x
\left[ \frac{1}{2}\sum_{\pm}\mp e^{\mp i\phi} \nabla_{z} \theta
(a^{\dagger}\sigma_{\pm} 
\stackrel{\leftrightarrow}{\nabla_{z}}a)
+\frac{1}{4}(\nabla_{z}\theta)^{2}(a^{\dagger}\sigma_{x}a)
\right] \nonumber\\
&=&\frac{\hbar^{2}}{2m}\sum_{{\bf k}}\sum_{q//z}
\left[ \sum_{\pm}\mp ie^{\mp i\phi}
 \left(k_{z}+\frac{q}{2}\right)A_{q}a^{\dagger}_{{\bf k}+q}
\sigma_{\pm}a_{{\bf k}}
+\frac{1}{4}\sum_{p//z}
A_{p}A_{-p+q}a^{\dagger}_{{\bf k}+q} a_{{\bf k}} \right].\label{Hint}
\end{eqnarray}
Here $(a^{\dagger}\stackrel{\leftrightarrow}{\nabla}a)\equiv 
a^{\dagger}({\nabla}a)-({\nabla}a^{\dagger})a$,
\begin{equation}
      A_{\bf q}\equiv  \frac{1}{V} \int d^{3}x e^{-iqz}\nabla_{z}\theta 
	=\frac{\pi}{L} e^{-iq z_{0}}\frac{1}{{\cosh}(\pi q\lambda/2)} 
	\delta_{{\bf q}_{\perp},0}
	\label{aqdef}
\end{equation}
is a classical gauge field which represents the domain wall, $V\equiv 
L_{\perp}^{2}L$ 
being the system volume and ${\bf q}\equiv ({\bf q}_{\perp},q)$, ${\bf 
q}_{\perp}$ being the momentum in $xy$-plane and $q$ is along $z$-direction. 
$A_{q}$ is finite only when ${\bf q}_{\perp}=0$, 
since we are considering a flat wall perpendicular to the $z$-axis.
Below we choose $\phi=\pi/2$.
We neglect the reaction to the wall from the conduction electron, 
since the wall is much heavier than the electron, and then the gauge 
field $a_{q}$ can be treated as a classical variable independent of 
time. 
Due to this gauge transformation, the electronic current in 
$z$-direction is 
changed to be
\begin{equation}
J_{z}\equiv -\int d^{3}x \frac{e\hbar}{m}\frac{i}{2} 
(c^{\dagger}\stackrel{\leftrightarrow}{\nabla}_{z}c)=J_{z}^{0}+\delta J,
	\label{current}
\end{equation}
where 
\begin{equation}
	J_{z}^{0}\equiv 
       \frac{e\hbar}{m} \sum_{{\bf k}} k_{z}a^{\dagger}_{{\bf k}} 
a_{{\bf k}}
	\label{j0def}
\end{equation}
is the usual current and there arises and additional current carried by the wall, 
\begin{equation}
\delta J\equiv  -\frac{e\hbar}{2m} \sum_{{\bf k},q//z}
A_{q}a^{\dagger}_{{\bf k}+q}\sigma_{x}a_{{\bf k}} .  \label{delJdef}
\end{equation}

\subsection{Disordered case}
We first consider a disordered case where the resistivity is dominated 
by normal impurities. The effect of the wall is calculated 
perturbatively based on Kubo formula. 
The conductivity $\sigma$ for the current along the wire is 
calculated from the imaginary
current-current correlation function 
\begin{equation}
Q(i\omega_{\ell})\equiv \frac{\hbar}{V} 
      <J_{z}(i\omega_{\ell})J_{z}(-i\omega_{\ell})>,
	\label{Qdef}
\end{equation}
where $\omega_{\ell}\equiv 2\pi\ell/\beta$ being the Matsubara 
frequency and $\beta\equiv 1/(k_{B}T)$) as (we define the Fourier 
transform of the electron as
$a_n\equiv (1/\sqrt{\beta}) \int_{0}^{\beta}d\tau e^{i\omega_{n}\tau} 
a(\tau)$) 
\begin{equation}
	\sigma=\lim_{\omega\rightarrow 0}{\rm Im}(Q(\hbar\omega+i0)-Q(i0))/\omega. 
	\label{sigmadef}
\end{equation}
Here $Q(\hbar \omega+i0)$ is the retarded correlation function 
obtained by analytical continuation, 
$Q(\hbar\omega+i0)\equiv Q(i\omega_{\ell} \rightarrow \hbar\omega+i0)$, 
and ${\rm Im}$ denotes the imaginary part.
We estimate the correction to the conductivity due to the wall 
to the second order of $A_{q}$. 
In this section we consider only the correction to the classical 
(Boltzmann) conductivity ($\sigma_{\rm c}$).
The quantum corrections represented by maximally crossed diagrams are 
considered in \S\ref{SECquantum}.

Zeroth order term of $Q$ is obtained as 
$Q_{0}(i\omega_{\ell})= -(e\hbar/m)^{2} (\hbar/V\beta)\sum_{n\kv \sigma}  
k_{z}^{2}G_{\kv n\sigma}G_{\kv,n+\ell,\sigma}$ and this leads to a 
well know classical conductivity due to the normal impurity, 
$\sigma_{0}\equiv e^{2}n\tau/m$ ($n$ being the electron density).
Here the imaginary time electron Green function includes the effect of the 
impurity and is given by
\begin{equation}
	G_{\kv,n,\sigma}\equiv
	 \frac{1}
	    {i(\omega_{n}+\frac{\hbar}{2\tau}{\rm sgn}(n))-\epsilon_{\kv\sigma} },
	\label{greenfunc}
\end{equation}
where $\omega_{n}\equiv\pi(2n-1)/\beta$ and ${\rm sgn}(n)=1$ and $-1$ for $n>0$ 
and $n<0$, respectively.
Green function carrying a frequency of $\omega_{n}+\omega_{\ell}$ is 
denoted by $G_{\kv,n+\ell,\sigma}$.

The first order contribution of $A_{q}$ vanishes.
The second order contributions to the Boltzmann conductivity
 are shown in Fig. \ref{FIGdiagram}.
The process $Q_{1}$ arises from the correction of the both of the two 
current 
vertices by the wall, $\delta J$, and $Q_{3}$ is due to the 
correction of 
one of the current vertex and a interaction with the wall.  
$Q_{2}$ and $Q_{4}$ are the self-energy due to the wall and $Q_{5}$ 
is the vertex correction.
These are written as 
\begin{eqnarray}
Q_{1}&=& 
-\left( \frac{e\hbar}{m} \right)^{2}\frac{1}{4}
\frac{1}{\beta}\sum_{n}\frac{1}{V}\sum_{\kv q\sigma}
|A_{q}|^{2}G_{\kv-\frac{q}{2},n,\sigma}
G_{\kv+\frac{q}{2},n+\ell,-\sigma}
\nonumber\\
Q_{2}&=& 
-\left( \frac{e\hbar}{m} \right)^{2}\frac{\hbar^{2}}{8m}\frac{1}{\beta}\sum_{n}
\frac{1}{V}\sum_{\kv q\sigma }
k_{z}^{2}|A_{q}|^{2}
[(
G_{\kv,n,\sigma})^{2}G_{\kv,n+\ell,\sigma}+
G_{\kv,n,\sigma}(G_{\kv,n+\ell,\sigma})^{2}
]
\nonumber\\
Q_{3}&=& 
-\left( \frac{e\hbar}{m} \right)^{2}\frac{\hbar^{2}}{2m}\frac{1}{\beta}\sum_{n}
\frac{1}{V}\sum_{\kv q\sigma}
k_{z}\left(k_{z}-\frac{q}{2}\right)|A_{q}|^{2}
[
G_{\kv-\frac{q}{2},n,\sigma}
G_{\kv-\frac{q}{2},n+\ell,\sigma}
G_{\kv+\frac{q}{2},n,-\sigma}
\nonumber\\
&&+
G_{\kv-\frac{q}{2},n,\sigma}
G_{\kv-\frac{q}{2},n+\ell,\sigma}
G_{\kv+\frac{q}{2},n+\ell,-\sigma}
]
\nonumber\\
Q_{4}&=& 
-\left( \frac{e\hbar}{m} \right)^{2}\frac{\hbar^{4}}{4m^{2}}\frac{1}{\beta}\sum_{n}
\frac{1}{V}\sum_{\kv q\sigma }
k_{z}^{2}\left(k_{z}-\frac{q}{2}\right)^{2} |A_{q}|^{2}
[(G_{\kv-\frac{q}{2},n,\sigma})^{2}
G_{\kv-\frac{q}{2},n+\ell,\sigma}
G_{\kv+\frac{q}{2},n,-\sigma}  
\nonumber\\
&& +
G_{\kv-\frac{q}{2},n,\sigma}
(G_{\kv-\frac{q}{2},n+\ell,\sigma})^{2}
G_{\kv+\frac{q}{2},n+\ell,-\sigma}]
\nonumber\\
Q_{5}&=& 
-\left( \frac{e\hbar}{m} \right)^{2}\frac{\hbar^{4}}{4m^{2}}\frac{1}{\beta}\sum_{n}
\frac{1}{V}\sum_{\kv q\sigma }
k_{z}^{2}\left(k_{z}^{2}-\frac{q^{2}}{4}\right) |A_{q}|^{2}
G_{\kv-\frac{q}{2},n,\sigma}
G_{\kv-\frac{q}{2},n+\ell,\sigma}
G_{\kv+\frac{q}{2},n,-\sigma}
G_{\kv+\frac{q}{2},n+\ell,-\sigma} .
\nonumber\\
&&
\end{eqnarray}

The contribution from the wall needs to vanish in the limit of 
vanishing Zeeman splitting, $\Delta\rightarrow 0$, since no 
scattering occurs. This is not 
easy to see by looking at each term, $Q_{i}$, but it will become 
obvious after we sum up these classical 
contributions to be $ Q_{1-5}\equiv \sum_{i=1,5}Q_{i}$.
In fact the summation can be carried out in a straight forward manner by use 
of identities
\begin{eqnarray}
G_{\kv,n,\sigma}G_{\kv,n+\ell,\sigma}&=&
-i{\left(
  \omega_{\ell}+\frac{\hbar}{2\tau}({\rm sgn}(n+\ell)-{\rm sgn}(n))
                \right)^{-1}}
	(G_{\kv,n,\sigma}-G_{\kv,n+\ell,\sigma}) 
	\nonumber\\
G_{\kv+\frac{q}{2},n,-\sigma} G_{\kv-\frac{q}{2},n,\sigma}
&=& \left(\frac{\hbar^{2} k_{z}q}{m}+2\sigma \Delta \right)^{-1}
 (G_{\kv+\frac{q}{2},n,-\sigma}-G_{\kv-\frac{q}{2},n,\sigma}).	
\end{eqnarray}
The result is (see Appendix \ref{APPQsum} for details of 
calculation) $Q_{1-5}=Q_{\rm c}+Q_{\rm c}'$, where
\begin{equation}
Q_{\rm c}(i\omega_{\ell})=
\frac{1}{2}\left(\frac{e\hbar\Delta}{m}\right)^{2}
\frac{1}{\beta}\sum_{n}\frac{1}{V}\sum_{\kv q\sigma }
|A_{q}|^{2}
G_{\kv-\frac{q}{2},n,\sigma}
G_{\kv-\frac{q}{2},n+\ell,\sigma}
G_{\kv+\frac{q}{2},n,-\sigma}
G_{\kv+\frac{q}{2},n+\ell,-\sigma}
,  \label{QSUMRESULT}
\end{equation}
and 
\begin{equation}
Q_{\rm c}'(i\omega_{\ell})=
-\frac{1}{4}\left(\frac{e\hbar}{m}\right)^{2}
\frac{1}{\beta}\sum_{n}\frac{1}{V}\sum_{\kv q\sigma }
|A_{q}|^{2} 
\Delta \frac{\Delta-\sigma\frac{(k_{z}+q/2)^{2}}{m}} 
{\left[ \Delta+\sigma\frac{(k_{z}+q/2)q}{2m} \right]^{2}} 
G_{\kv,n,\sigma} G_{\kv,n+\ell,\sigma}
.  \label{Qp}
\end{equation}
The term $Q_{\rm c}$ is dominant and the 
contribution from the term $Q_{\rm c}'$ turns out to cancel 
with the effect of the shift of the electron density, which is 
calculated later.

The summation over Matsubara frequency, $\omega_{n}$, in eqs. (\ref{QSUMRESULT}) 
and (\ref{Qp}) can be carried 
out by use of contour integration (see Appendix \ref{APPomegasum})
and the contribution to the Boltzmann conductivity from the five 
classical processes,
$\sigma_{1-5}\equiv 
\lim_{\omega\rightarrow 0}{\rm Im}( Q_{1-5}(\omega+i0)- Q_{\rm 
c}(i0))/\omega$, 
is obtained as $\sigma_{1-5}=\sigma_{\rm c}+\sigma_{\rm c}'$, where
\begin{equation}
\sigma_{\rm c}=
-\frac{\Delta^{2}\hbar^{3}}{8\pi\tau^{2}}\left(\frac{e\hbar}{m}\right)^{2}
\frac{1}{V}\sum_{\kv q\sigma }|A_{q}|^{2}
\frac{(\epsilon_{\kv-\frac{q}{2},\sigma}
       +\epsilon_{\kv+\frac{q}{2},-\sigma})^{2}}   
   {\left[(\epsilon_{\kv-\frac{q}{2},\sigma})^{2}
         +\left(\frac{\hbar}{2\tau}\right)^{2}\right]^{2}
    \left[(\epsilon_{\kv+\frac{q}{2},-\sigma})^{2}
         +\left(\frac{\hbar}{2\tau}\right)^{2}\right]^{2}},
         \label{sigmacresult}
\end{equation}
and $\sigma_{\rm c}'$ is the contribution from $Q_{\rm c}'$;
\begin{equation}
\sigma_{\rm c}'= \frac{\hbar}{32\pi} \left( 
\frac{e\hbar}{m}\right)^{2} \frac{\Delta}{\tau^{2}}
\frac{1}{V}\sum_{\kv q\sigma} 
\frac{|A_{q}|^{2}}
 {\left[\epsilon_{\kv\sigma}^{2}+\left(\frac{\hbar}{2\tau}\right)^{2}\right]^{2}}
 \frac{\Delta-\sigma\frac{(k_{z}+q/2)^{2}}{m}} 
{ \left[\Delta+\sigma\frac{(k_{z}+q/2)q}{2m} \right]^{2}}.
	\label{sigmap}
\end{equation}

Besides the processes in Fig. \ref{FIGdiagram}, 
there is another contribution the classical conductivity, which is due 
to the change of the electron density in the presence of a domain wall.
The correction to the electron density due to the interaction 
(\ref{Hint}) is written as (diagramatically shown in Fig. 
\ref{FIGdeltan})
\begin{equation}
	\delta n=\frac{\hbar^{2}}{4m}\frac{1}{\beta V}\sum_{\kv q n\sigma} 
|A_{q}|^{2} \left[ \frac{1}{2}(G_{\kv n\sigma})^{2} 
+\frac{\hbar^{2} k_{z}^{2}}{m}
 (G_{\kv-\frac{q}{2}, n\sigma})^{2} G_{\kv+\frac{q}{2}, n,-\sigma} 
 \right].
	\label{delndef}
\end{equation}
After some calculation it reduces to (Appendix \ref{APPdeln})
\begin{equation}
\delta n= -\frac{\hbar^{3}}{16\pi m\tau }\frac{1}{V}\sum_{\kv q\sigma} 
\frac{|A_{q}|^{2}}
 {\epsilon_{\kv\sigma}^{2}+\left(\frac{\hbar}{2\tau}\right)^{2}}
 \frac{\Delta-\sigma\frac{(k_{z}+q/2)k_{z}}{m}} 
{\Delta+\sigma\frac{(k_{z}+q/2)q}{2m} }.
	\label{deln}
\end{equation}
this shift of the electron density leads to a correction of the 
zeroth order conductivity, $\sigma_{0}\rightarrow 
\sigma_{0}+\delta\sigma_{\rm c}$, where $\delta\sigma_{\rm c}=e^{2}\tau\delta n /m$ 
is obtained from eq. (\ref{deln}) as
\begin{equation}
\delta \sigma_{\rm c}= -\frac{\hbar}{16\pi} \left( 
\frac{e\hbar}{m}\right)^{2} \frac{1}{V}\sum_{\kv q\sigma} 
\frac{|A_{q}|^{2}}
 {\epsilon_{\kv\sigma}^{2}+\left(\frac{\hbar}{2\tau}\right)^{2}}
 \frac{\Delta-\sigma\frac{(k_{z}+q/2)k_{z}}{m}} 
{\Delta+\sigma\frac{(k_{z}+q/2)q}{2m} }.
	\label{delsigma}
\end{equation}
It turns out after the $\kv$-summation that 
$\sigma_{\rm c}'+\delta\sigma_{\rm c}$ vanishes in the case of 
$k_{F}\lambda\gg1$ 
(Appendix\ref{APPdeln}).
Thus the classical correction to the conductivity due to the wall is 
given by $\sigma_{\rm c}$.

To proceed further we neglect quantities of $O((q/k_{F})^{2})$ and 
approximate $\epsilon_{ \kv\pm q/2,\mp\sigma}\simeq 
\epsilon_{\kv}\pm[(\hbar^{2}  k_{z}q/2m)+\sigma\Delta]$.
This is because the momentum transfer, $q$, is limited to a small value of 
$q\lesssim \lambda^{-1}$ due to the 
form factor of the wall, $|A_{q}|^{2}\propto [\cosh(\pi q\lambda/2)]^{-2}$, and we are considering the 
case of a thick wall, $k_{F}\lambda \gg 1$.
The result of $\kv$-summation is (see Appendix \ref{APPksum}) 
\begin{equation}
	  \sigma_{\rm c}=
	-\frac{e^{2}\Delta^{2}\tau^{2}}{8\pi\hbar^{3}}n_{\rm w}\sum_{\sigma} 
  \int_{-\infty}^{\infty}\frac{dx}{x}\frac{1}{\cosh^{2}x}
\left[ 
\tan^{-1} 
\left(\frac{2l_{\sigma}}{\pi \lambda}x+ 2\Delta
\frac{\tau}{\hbar}\right)
+
\tan^{-1} 
\left(\frac{2l_{\sigma}}{\pi \lambda}x-2\Delta
\frac{\tau}{\hbar}\right)
\right],
	\label{SIGMAC}
\end{equation}
where $x\equiv \pi q \lambda/2$ and
$l_{\sigma}\equiv \hbar k_{F\sigma}\tau/m$ is the mean free path of the 
electron with spin $\sigma$,
$n_{\rm w}\equiv 1/L$ being the density of the wall.

Using this result the contribution of the wall to the Boltzmann resistivity is 
given  as
\begin{equation}
	\rho_{\rm c}\equiv (\sigma_{0}+\sigma_{\rm c})^{-1} -\sigma_{0}^{-1} \simeq 
	\frac{|\sigma_{\rm c}|}{\sigma_{0}^{2}}.
	\label{rhocdef}
\end{equation}
The last expression is valid if the contribution from the impurity is much 
larger than that from the wall, namely if $|\sigma_{\rm c}|/\sigma_{0} \ll 1$.
(But see also the discussion below eq. (\ref{rhow}))

Here we consider the case of a ferromagnet with weak disorder, and 
assume two further conditions; 
\begin{equation}
\Delta\tau/\hbar \gg1,  \label{condition1}
\end{equation}
which indicates that the effect of Zeeman splitting is not smeared by 
the width of energy level, and 
\begin{equation}
m\Delta\lambda/k_{F}\hbar^{2} \gg 1. \label{condition2}
\end{equation}
The second condition is satisfied if the Zeeman splitting is not too 
small. 
Both inequalities would be satisfied in the case of $d$ electron. 
In this case the classical correction due to the wall is obtained as 
(eq. (\ref{sigmacferro}))
\begin{equation}
	\sigma_{\rm  c} \simeq -\frac{e^{2}}{4\pi^{2}\hbar} n_{\rm w}
	\sum_{\sigma}
	\frac{l_{\sigma}}{\lambda} \int_{0}^{\infty} \frac{dx}{\cosh^{2}x}
	= -\frac{e^{2}}{4\pi^{2}\hbar}n_{\rm w}\sum_{\sigma}
	\frac{l_{\sigma}}{\lambda}
		\;\;\;\;\;\;\;\;\;\;\;(\Delta\tau/\hbar\gg 1, m\Delta\lambda/k_{F}\hbar^{2} \gg 1).
	\label{SIGMACFERRO0}
\end{equation}

\subsection{Ballistic limit}
\label{APPclean}

The ballistic limit, $l\rightarrow\infty$,  is also of recent interest, as 
a large magnetoresistance is expected there as is indeed
realized in nano scale contacts\cite{Garcia99}.
In this limit
the perturbative calculation with respect to domain wall contribution 
relative to the normal impurity scattering based on Kubo formula fails.
In this case Mori formula becomes valid, which relates the resistivity 
with the correlation of random forces as\cite{Mori65}
\begin{equation}
\rho_{\rm c}=\left(\frac{e^{2} n}{m}\right)^{-2}\lim_{\omega\rightarrow 0}
\frac{1}{\hbar\omega} 
{\rm Im}[\chi_{\dot{J}\dot{J}}(\hbar\omega)-\chi_{\dot{J}\dot{J}}(0)].
\label{Mori}
\end{equation}
Here $\chi_{\dot{J}\dot{J}}(i\omega_{\ell})\equiv
-(\hbar/V) <\dot{J}(i\omega_{\ell})\dot{J}(-i\omega_{\ell})>$
with $\dot{J}\equiv dJ/dt= \frac{ i}{\hbar}[H,J]$.
$\omega_{\ell}\equiv 2\pi\ell/\beta$ is a Bosonic Matsubara 
frequency ($\beta\equiv 1/(k_{B}T)$).
The correlation function 
$\chi_{\dot{J}\dot{J}}(\hbar\omega)$ in eq. (\ref{Mori}) denotes 
an analytical continuation of the correlation function calculated for 
imaginary-time frequency, 
i.e., $\chi_{\dot{J}\dot{J}}(\hbar\omega)\equiv 
\chi_{\dot{J}\dot{J}}(i\omega_{\ell}\rightarrow \hbar \omega+i0)$.

The non-conservation of the current (i.e., finite $\dot{J}$) arises from the 
scattering by the wall. 
In fact eq. (\ref{Hint}) and (\ref{delJdef}) lead to 
\begin{equation}
\dot{J_{z}}\equiv \frac{ i}{\hbar}[H,J_{z}]=i\left(\frac{e }{m}\right) \Delta
\sum_{\kv q\sigma }\sigma
A_{q} a^{\dagger}_{\kv+q}\sigma_{x} a_{\kv} .
\end{equation}
The imaginary-time correlation function is then calculated as
\begin{equation}
\chi_{\dot{J}\dot{J}}(i\omega_{\ell}) =
-\frac{\hbar}{L}\left(\frac{e\Delta}{m}\right)^{2} \sum_{kq\sigma} |A_{q}|^{2}
\frac{1}{\beta}\sum_{n}G_{k+q,n+\ell,-\sigma} G_{kn\sigma}.
\end{equation}
The summation over $\omega_{n}$ is carried out to obtain
\begin{equation}
\chi_{\dot{J}\dot{J}}(i\omega_{\ell})=
-\frac{\hbar}{L}\left(\frac{e\Delta}{m}\right)^{2} \sum_{kq\sigma} |A_{q}|^{2}
\frac{
  f(\epsilon_{k+q,-\sigma})- f(\epsilon_{k,\sigma})  }
  { \epsilon_{k+q,-\sigma} -\epsilon_{k,\sigma} -i\omega_{\ell} },
	\label{chidotcor}
\end{equation}
Thus the resistivity is calculated as
\begin{eqnarray}
\rho_{\rm c}&=& \frac{\pi \Delta^{2}}{e^{2}n^{2}}\frac{1}{V}
 \sum_{\kv q \sigma} |A_{q}|^{2} 
\delta(\epsilon_{\kv+q,-\sigma} -\epsilon_{\kv,\sigma} ) \delta(\epsilon_{\kv,\sigma} )
\nonumber\\
&\simeq& \frac{\pi^{2} \Delta^{2}}{2e^{2}n^{2}V} n_{\rm w} \sum_{\sigma}N_{\sigma}
 \int_{-\infty}^{\infty}\frac{dq}{\cosh^{2}\frac{\pi}{2}q\lambda} 
 \int_{-1}^{1}\frac{d\cos\theta}{2} 
 \delta\left(\frac{\hbar k_{F\sigma}q}{m}\cos\theta+2\sigma\Delta \right)
 \nonumber\\
&=&
\frac{m^{2} \Delta^{2}}{4e^{2}n^{2}\hbar^{3}}n_{\rm w} \sum_{\sigma}
 \int_{\Lambda_{c}(\sigma)}^{\infty} \frac{dx}{x} \frac{1}{\cosh^{2}x} ,
\label{rhocclean}
\end{eqnarray}
where $\Lambda_{c}(\sigma)\equiv (\pi m 
\Delta\lambda/k_{F\sigma}\hbar^{2})$ and $x\equiv \pi q\lambda/2$.
In bulk ferromagnet with thick wall ($\lambda \gg k_{F}^{-1} $) and not 
very small Zeeman splitting, $\Lambda_\sigma$ is large and thus the 
resistivity due to the wall is exponentially small,
\begin{equation}
	\rho_{\rm c}\simeq \frac{m^{2}\Delta^{2}n_{\rm w} }{2e^{2}n^{2}\hbar^{3}} 
	\sum_{\sigma}
	\frac{e^{-2\Lambda_\sigma}}{\Lambda_\sigma}
	\;\;\;\;\;\;\;\;\;\;\;(\tau\rightarrow\infty, \Lambda_\sigma \gg 1). 
	\label{cleanlimitrho}
\end{equation}
This is because the electron spin can follow adiabatically the rotation of 
the magnetization as it passes through the wall and so there will be very 
small reflection\cite{Cabrera74}.

On the other hand in nanoscale contacts where the thickness of the wall can 
be determined mostly by the sample geometry, the wall can be very 
thin\cite{TG99}. In this limit, $\lambda_{c}\rightarrow0$, we 
obtain  
\begin{equation}
	\rho_{\rm c}\simeq \frac{m^{2}\Delta^{2}n_{\rm w} }{4e^{2}n^{2}\hbar^{3}} 
	\sum_{\sigma}
	|\ln\Lambda_\sigma|
	\;\;\;\;\;\;\;\;\;\;\;(\tau\rightarrow\infty, \Lambda_\sigma 
	\rightarrow0). 
\end{equation}
Thus the resistance due to the wall, $R_{\rm c}\equiv \rho_{\rm 
c}L/A_{\perp}$ in this case is obtained as
\begin{equation}
	R_{\rm c}= \frac{h}{e^{2}} \frac{9\pi^3}{32} 
	\frac{(k_{F+}^2-k_{F-}^2)^2}{(k_{F+}^3+k_{F-}^3)^2 A_{\perp}}
    \left| \ln \left(\frac{\pi}{4}\right)^2 
    \frac{(k_{F+}^2-k_{F-}^2)^2 \lambda^2}{k_{F+}k_{F-}}\right|.
\end{equation}
It is seen that in the case of atomic size contact the resistance 
due to a domain wall can be of order of $h/e^2$, and thus a 
large magnetoresistance is expected by switching the magnetization 
direction in a atomic size contact, where the conductance in the 
absence of the wall is of order of quantized conductance, 
$e^2/h$\cite{TG99}.
In contrast in disordered case the effect in the limit of thin wall 
is not as large as in the ballistic case\cite{TG99}. 

It is interesting that the result of the resistivity based on Kubo formula, 
eq. (\ref{rhocdef})  with $\sigma_{\rm c}$ given by (\ref{SIGMAC}), 
has the identical limiting value at $\tau\rightarrow \infty$ (the 
first term in eq. (\ref{sigmac3})) as the result of Mori formula 
(eq. (\ref{rhocclean})), which suggests that the Kubo formula 
calculation remains correct in this limit, 
although the perturbative treatment used there becomes invalid.
Furthermore it turns out that if we assume finite lifetime for the electron 
in Mori formula, which is again without justification,
the result is identical to the result of Kubo formula (\ref{rhocdef}).
 In fact Mori formula treated this way leads to 
\begin{equation}
\rho_{\rm c}=
\left({e n}\right)^{-2}\frac{\hbar^{3}}{4\pi}
\left(\frac{\hbar \Delta}{\tau}\right)^{2}
\frac{1}{V}\sum_{\kv q\sigma }|A_{q}|^{2}
\frac{1}
{\left[\epsilon_{\kv-\frac{q}{2},\sigma}^{2}
         +\left(\frac{\hbar}{2\tau}\right)^{2}\right]
    \left[\epsilon_{\kv+\frac{q}{2},-\sigma}^{2}
         +\left(\frac{\hbar}{2\tau}\right)^{2}\right]},
         \label{rhow}
\end{equation}
and this is shown to be equivallent to eqs.  (\ref{SIGMAC}) and 
(\ref{rhocdef}) by similar calculation as in \S\ref{APPksum}.
These facts  may indicate the correctness of the expression (\ref{SIGMAC}) and 
(\ref{rhocdef}) for any value of $\tau$.

%
\section{Quantum Correction}
\label{SECquantum}

In the clean case, where $k_F \ell$ is large, the classical 
theory of transport considered in the previous section describes
the electronic resistivity well.
In disordered case with smaller $k_{F} \ell$, a correction becomes  
necessary, since the quantum nature of the electron begins to be 
important as a result of strong scattering by elastic impurities. 
Elastic impurities are those which cause only elastic scattering and 
thus do not destroy the coherence of the electron.
In this case the scattered 
electron wave interferes with the incoming wave, and a state like a 
standing wave is formed. 
In this weakly localized state the conductivity is suppressed due to 
the quantum interference.
The effect is stronger in lower dimensions, where the backward scattering 
occurs more effectively.
This weakly localized state is described conveniently in terms of an 
amplitude of two electrons propagating in opposite direction  
interacting with each other through multiple impurity scatterings
 (Fig. \ref{FIGCooperon}).  
This process of particle-particle propagation is called Cooperon.
The amplitude is related to the amplitude of the backward scattering, 
and is shown to grow at long wavelength because of the quantum 
interference.
In fact Cooperon for the two incoming electrons with 
the same spin $\sigma$ and the momentum of $k_{F}$ and 
$-k_{F}+q$ is calculated at small $q$ and small thermal frequencies 
($\omega_n$ and $\omega_{n'}$) as\cite{Lee85}  
\begin{equation}
\Gamma_{+-}^{\sigma\sigma}(q) \simeq 
\frac{ n_{\rm i}v^{2} }{(Dq^{2}+\omega_n -\omega_{n'})\tau}
\equiv \Gamma_{0}(q,n-n') ,
\label{cooperon1}
\end{equation} 
where $D\equiv (\hbar^{2} k_{F}^{2}\tau/3m^{2})$ and $_{+-}$ denotes 
$\omega_{n}>0$ and $\omega_{n'}<0$ (See Appendix \ref{APPCooperon}). 
(In calculating the quantum correction in this section we neglect difference 
between two $k_{F\sigma}$'s, i.e., quantity of $O(\Delta/\epsilon_{F})$ 
for simplicity.)
We consider here a ferromagnet with 
$\Delta \tau /\hbar\gg1$ and then 
Cooperons connecting the electrons with opposite 
spin does not lead to any singular contribution  
and is thus neglected (Appendix \ref{APPCooperon}).

In (\ref{cooperon1}) we have neglected the effect of inelastic scattering.
In reality, the singularity of eq. (\ref{cooperon1}) is 
smeared out due to the inelastic scattering (e.g., by phonons,
electron-electron interaction and spin flip scattering by magnetic 
impurities) at finite temperature. 
Here we treat these inelastic scattering phenomenologically by
introducing an inelastic lifetime, $\tau_\varphi$ 
(see Refs. \cite{Bergmann84,Lee85} for microscopic discussion).
A finite value of $\tau_\varphi$ indicates that the coherence is kept 
only to the spatial scale of $\ell_{\varphi}\equiv 
\sqrt{D\tau_{\varphi}}$.
The Cooperon is modified then to be 
\begin{equation}
 \Gamma_{0}(q,n-n')=
 \frac{ n_{\rm i}v^{2} }
 {(Dq^{2}+\omega_n -\omega_{n'}+1/\tau_{\varphi})\tau} .
\label{cooperon2}
\end{equation} 
In the calculation of the static conductivity the 
thermal frequencies in the Cooperon ($\omega_n$ and $\omega_{n'}$) can 
be set equal to zero, and thus we write 
$\Gamma_{0}(q)\equiv \Gamma_{0}(q,\omega_n=0)$ in this section.
In terms of the Cooperon the quantum correction to the conductivity 
is written in the absence of the wall as (eq. (\ref{sigma0q}))
\begin{equation}
	\sigma_{0{\rm q}}=-\frac{1}{2\pi} \left(\frac{e\hbar}{m}\right)^2 
	\frac{1}{V} \frac{4\pi\hbar k_{F}^2}{3} N(0) 
	\left(\frac{\tau}{\hbar}\right)^3 \sum_{\qv,\sigma}
	\Gamma_{0}(q),
	\label{sigma0q_1}
\end{equation}
which is diagramatically expressed as in Fig. \ref{FIGsigma0q}.
This diagram is called a maximally crossed diagram, since the lines 
denoting the impurity scattering is crossing each other maximally.
The quantum correction is rewritten as
\begin{eqnarray}
\sigma_{0{\rm q}} &\simeq & - 2\sigma_{0}\left(\frac{\tau}{\hbar}\right)^{2} 
\sum_{\qv} \Gamma_{0}(q)
\nonumber\\
&=& - 2\pi \sigma_{0}\left(\frac{\tau}{\hbar}\right) 
\frac{\hbar^2}{mk_{F}}
\frac{1}{V} \sum_{\qv} \frac{1}{Dq^2\tau+\kappa_{\varphi}},
	\label{Qq01}
\end{eqnarray}
where $\kappa_{\varphi}\equiv \tau/\tau_{\varphi}$ and we have used 
$2\pi n_{\rm i}v^{2}N(0)\tau/\hbar=1$.
The summation over $q$ is cut off for small $q$ at $q_{z}\sim \pi L^{-1}$ 
in the 
wire direction and $q_{\perp}\sim \pi L_{\perp}^{-1}$ in the perpendicular 
direction.
Here we consider the case where the perpendicular dimension of the 
wire is small as compared with inelastic diffusion length, 
namely $L_{\perp} \lesssim \ell_{\varphi}$. 
In this case $q$-summation can be carried out in one-dimension along 
$z$-direction and we obtain
\begin{equation}
	\frac{1}{V}\sum_{\qv} \frac{1}{Dq^2\tau+\kappa_{\varphi}}=
	\frac{1}{L_{\perp}^2} \frac{1}{\pi} \int_{\pi/L}^{\pi/\ell} dq
	\frac{1}{(q\ell)^2/3+\kappa}
	\simeq \frac{3\ell_{\varphi}}{2 L_{\perp}^2 \ell^2} ,
	 \label{qsumcooperon}
\end{equation}
where we have used $\kappa \ll 1$. In the last equality we used 
$L\gg \ell_{\varphi}$, which we assume. 
The result of the conductivity correction is thus  
\begin{equation}
\sigma_{0{\rm q}} 
\simeq -3\pi \sigma_{0} \frac{\ell_{\varphi}}{k_{F}^2 \ell 
L_{\perp}^2}.
\label{sigma0qresult}
\end{equation}
\subsection{Dephasing effect due to domain wall}
Now we calculated the effect of the domain wall on the quantum 
correction.
The quantum correction arises when the coherence of the electron 
system is modified by the wall.
There is thus no effect if the magnetization change inside the wall is 
slow enough, in which case the electron can follow the magnetization 
change easily. 
The decoherence occurs due to a non-adiabatic process.
In fact the first term in the interaction with the wall, eq. 
(\ref{Hint}), can cause dephasing since it flips the electron spin. 
Thus the interference effect among electrons is suppressed due to 
this interaction with the wall.
This dephasing effect is most conveniently studied in terms of the 
correction to the Cooperon.
The correction to the Cooperon by the wall is shown up to the second 
order in Fig. \ref{FIGQC2nd}.
The process (a) is the most dominant contribution.
Diagram (b) contributes to dephasing but 
is not important since it contains only one Cooperon and 
thus gives smaller contribution compared to process (a).
Similar diagram (c) and the vertex correction type (d)(e) are neglected 
because they include Cooperons with different spins and thus are 
suppressed by a factor of $O(\hbar/\Delta\tau)$.
The effect of dephasing due to the process (a) 
becomes clear by summing up higher order processes shown in 
Fig. \ref{FIGCooperonDW}, which gives rise to a mass of Cooperon (see 
eq. (\ref{Cooperonwall})).
The Cooperon in the presence of the wall is calculated from these 
consideration as
\begin{equation}
\Gamma_{\rm w} \equiv \Gamma_{0}+
\Gamma_{0}^{2 }I_{\rm w} +\Gamma_{0}^{3 }I_{\rm w}^{2} 
     \cdots \simeq 
	\Gamma_{0}\left[ \frac{1}{1-\Gamma_{0}I_{\rm w}} \right],
	\label{gammaDWdef}
\end{equation}
where
\begin{equation}
	I_{\rm w}\equiv \sum_{\kv'} (G_{\kv',n,\sigma})^{2} G_{-\kv'+\qv,n',\sigma} 
	\Sigma^{\rm w}_{\kv',n,\sigma} +{\rm c.c.}, \label{Iwdef}
\end{equation}
and 
\begin{equation}
	\Sigma^{\rm w}_{\kv',n,\sigma}  \equiv \left( \frac{\hbar^{2}}{2m} \right)^{2} 
	\sum_{q//z} \left( k_{z}'+\frac{q}{2}\right) ^{2} |A_{q}|^{2} 
	G_{\kv'+q,n,-\sigma},
	\label{selfenergy}
\end{equation}
is the self-energy due to the wall scattering.
For the case of $k_{F}\lambda \gg 1$ and $\Delta\tau/\hbar \gg1$ we are 
considering, this is evaluated in the case of electron at the Fermi 
energy ($|\kv'|=k_{F}$) as (we consider the case $\omega_n >0$)
\begin{eqnarray}
\Sigma^{\rm w}_{\kv',n,\sigma}  
&=& 
\left(\frac{\hbar^{2}}{2m} \right)^{2}
\frac{1}{\frac{i\hbar}{2\tau}-\epsilon_{\kv'+q,-\sigma}}
\sum_{q//z} \left(k_{F}\cos\theta+\frac{q}{2}\right)^{2} |A_{q}|^2
\nonumber\\
& \simeq & \left(\frac{\hbar^{2}k_{F}\cos\theta}{2m} \right)^{2}
\frac{1}{ \frac{i\hbar}{2\tau}-2\sigma\Delta }
\sum_{q//z} |A_{q}|^2
\nonumber\\
& = & 2\left( \frac{\hbar^{2}k_{F}\cos\theta}{2m} \right)^{2}
\frac{1}{ \frac{i\hbar}{2\tau}-2\sigma\Delta } \frac{n_{\rm w}}{\lambda},
\end{eqnarray}
where $\theta$ is the angle of the incoming momentum $\kv'$ measured from the 
$z$-axis.

Its imaginary part of $\Sigma^{\rm w}$ is related to the life-time due to the 
scattering by the wall as 
\begin{eqnarray}
	\frac{1}{\tau_{\rm w}(\theta,\sigma)} &\equiv& 	
	-\frac{2}{\hbar} {\rm Im} \Sigma^{\rm w}_{\kv',n>0,\sigma} 
	\nonumber\\
	&=& \frac{1}{2}\frac{n_{\rm w}}{k_{F}^2 \lambda} 
	\left(\frac{\epsilon_{F}\cos\theta}{\Delta} \right)^2 \frac{1}{\tau} .
	\label{tauwdef}
\end{eqnarray}
The real part is neglected since it has only irrelevant 
effect of shifting the energy. 
By use of $\tau_{\rm w}$, $I_{\rm w}$ of (\ref{Iwdef}) is calculated as
\begin{eqnarray}
I_{\rm w} & \simeq& -\frac{i}{2}N(0) \int_{-\infty}^{\infty} d\epsilon 
\frac{1}{\left( \frac{i\hbar}{2\tau}-\epsilon\right)^{2} 
        \left( -\frac{i\hbar}{2\tau}-\epsilon\right) } 
        \left< \frac{1}{\tau_{\rm w}(\theta)} \right>+{\rm c.c.}
        \nonumber\\
        &=&
        -2\pi N(0) \kappa_{\rm w},
\end{eqnarray}
where brackets denotes the average over $\theta$ and 
$\kappa_{\rm w}$ is given as 
\begin{equation}
\kappa_{\rm w}\equiv\left< \frac{\tau}{\tau_{\rm w}} \right>= 
\frac{ n_{\rm w}}{6\lambda k_{F}^{2}} 
\left(\frac{\epsilon_{F}}{\Delta}\right)^{2}  .
	\label{kappaw}
\end{equation}
We therefore obtain the Cooperon in the presence of the wall (eq. 
(\ref{gammaDWdef})) as
\begin{equation}
\Gamma_{\rm w}(\qv)=
 \frac{n_{\rm i}v^{2} }{Dq^{2}\tau+\kappa _{\varphi}+\kappa _{\rm w} } .
	\label{Cooperonwall}
\end{equation}
It is seen that the spin flip scattering by the wall contributed to 
an additional dephasing time. 
Note that the electron in the Cooperon here refers to the electron 
after the gauge transformation (\ref{gaugetr}) , i.e., $a$.

In the presence of the wall, the quantum corrections given by  
eq. (\ref{Qq01})) but with $\Gamma_0$ replaced by $\Gamma_{\rm w}$.
Thus the contribution of the wall to the quantum correction is given 
by (by use of (\ref{Qq01}))
\begin{equation}
	\delta\sigma_{{\rm q}} = 
	 2\sigma_{0}\left(\frac{\tau}{\hbar}\right)^{2} 
\sum_{\qv} (\Gamma_{0}(q)-\Gamma_{\rm w}).
\end{equation}
In the same way as eq. (\ref{sigma0qresult}), we obtain
\begin{equation}
\delta\sigma_{{\rm q}} =
 \sigma_{0} \frac{\sqrt{3} \pi}{ k_{F}^2 L_{\perp}^2 } 
\left(\frac{1}{ \sqrt{\kappa_{\varphi}} }
    -\frac{1}{ \sqrt{ \kappa_{\varphi}+\kappa_{\rm w} } } \right).
    \label{sigmaqwall}
\end{equation}
This quantum correction is positive, since the wall suppresses the 
localization by causing the dephasing.

\section{Total resistivity due to domain wall}

We have calculated both the classical and quantum contribution to the 
conductivity from the wall. 
The full conductivity due to the wall is obtained as 
$\Delta\sigma_{\rm w}\equiv \sigma_{\rm c}+\delta\sigma_{\rm q}$ from 
eqs. (\ref{SIGMAC}) and (\ref{sigmaqwall}).  
In the clean limit there is no quantum correction.
In the case of the disordered case
with conditions (\ref{condition1}) and (\ref{condition2}) satisfied, 
the full correction to the conductivity by the wall is obtained as 
(by use of $\sigma_{0}=\frac{e^2}{\hbar}\frac{k_{F}^2 \ell}{3\pi^2}$) 
\begin{equation}
\Delta\sigma_{\rm w}=\sigma_{0}
\left[ -\frac{3}{2}\frac{n_{\rm w}}{k_{F}^2 \lambda}
+\frac{\sqrt{3}\pi}{(k_{F} L_{\perp})^2} 
 \left( \frac{1}{\sqrt{\kappa_{\varphi}}} 
    - \frac{1}{\sqrt{\kappa_{\varphi}+\kappa_{\rm w}}}  \right) 
    \right],
	\label{Deltasigmawall}
\end{equation}
where we have neglected $o(\Delta/\epsilon_{F})$.
The conductivity in the absence of the wall is expressed including 
the quantum correction ($\sigma_{0{\rm q}}$) as
\begin{equation}
\sigma=\sigma_{0}
\left[ 1-\frac{\sqrt{3}\pi}{(k_{F} L_{\perp})^2} 
\frac{1}{\sqrt{\kappa_{\varphi}}} 
    \right].
	\label{fullsigmanowall}
\end{equation}
In terms of the resistivity, which is the reciprocal of the 
conductivity, the wall contribution is obtained as 
$\Delta\rho_{\rm w}\equiv (\sigma+\Delta\sigma_{\rm w})^{-1}-\sigma^{-1}
\simeq -\Delta\sigma_{\rm w}/\sigma^2$.
The relative ratio of the wall contribution and the full 
resistivity ($\rho\equiv \sigma^{-1}$) is thus obtained as
\begin{equation}
\frac{\Delta\rho_{\rm w}}{\rho} \simeq
 \frac{3}{2}\frac{n_{\rm w}}{k_{F}^2 \lambda}
       -\frac{\sqrt{3}\pi}{(k_{F} L_{\perp})^2} 
 \left( \frac{1}{\sqrt{\kappa_{\varphi}}} 
    - \frac{1}{\sqrt{\kappa_{\varphi}+\kappa_{\rm w}}}  \right)  
,
	\label{rhowratio}
\end{equation}
where we have neglected the quantum correction in $\rho^{-1}$ (eq. 
(\ref{fullsigmanowall})).
The first term is the classical contribution and the 
second term is the quantum contribution, which reduces the resistivity.
The classical contributions $n_{\rm w}/(k_{F}^2\lambda)$ is 
proportional to the reflection probability 
caused by the interaction (\ref{Hint}), which is not large in 
conventional 3$d$ transition metals where $k_{F}\lambda$ is large.
The quantum contribution is also proportional to this probability 
(it is essentially $\tau_{\rm w}^{-1}$ of eq. (\ref{kappaw})), 
but in this case the 
effect is enhanced at by the coherence, as indicated by a large factor 
of $1/\sqrt{\kappa_{\varphi}}$.  
Thus at lower temperature where $\kappa_{\varphi}$ is smaller 
the quantum dephasing effect wins over the classical effect and thus the 
resistivity contribution from the wall can be negative (see below).

In experiments the effect of dephasing by the wall would be separated 
from other classical effects by looking into the temperature 
dependence. In fact the dephasing due to phonons, electron-electron 
interaction is known to decrease at lower temperature, typically 
by a power law, $\tau_{\varphi}^{-1} \propto T^\alpha$, $\alpha$ being 
a constant of $O(1)$\cite{Bergmann84,Lee85}. Thus according to the result 
(\ref{rhowratio}) the decrease of the resistivity due to the 
dephasing effect will increase as 
\begin{equation} 
\frac{\Delta\rho_{\rm w}}{\rho}\propto T^{-3\alpha/2}, \label{rhowtdep}
\end{equation}
in the region $\tau_{\rm w}^{-1} \ll \tau_{\varphi}^{-1}$.

It has been discussed that at low temperature, e.g. below $1$K, the 
dephasing time $\tau_{\varphi}$ is mostly due to the 
electron-electron interaction\cite{Bergmann84,Lee85}. The temperature 
dependence of $\tau_{\varphi}$ in this case has been estimated to be
$\tau_{\varphi}^{-1}\simeq [(k_{B}T/\sqrt{\tau})/(k_{F}L_{\perp})^2]^{2/3}$ 
in wires, and thus $\alpha=2/3$\cite{Altshuler82}. 
Let us consider a wire of $L_{\perp}=150$\AA\ ($K_{F}L_{\perp}=100$, 
$k_{F}^{-1}\simeq 1.5$\AA). If the elastic mean free path is 
$\ell\sim30$\AA, which corresponds to $\epsilon_{F}\tau/\hbar\sim 10$, 
the above expression leads to 
$\kappa_{\varphi}=\tau/\tau_{\varphi} \simeq 0.22\times 10^{-4}$ at 
$T=1$K. 
The dephasing effect due to the wall is calculated from eq. 
(\ref{kappaw}). Considering a wire length of $L=n_{\rm 
w}^{-1}=1000$\AA\ and a thin wall $k_{F}\lambda=200$ as observed in Co 
film\cite{Gregg96} and choosing $\Delta/\epsilon=0.2$, we obtain 
$\kappa_{\rm w}=3.1\times 10^{-5}$. The relative contribution from the 
wall $\Delta\rho_{\rm w}/{\rho}$ in this case is then calculated as
$\Delta\rho_{\rm w}/{\rho}=-0.051$. Note that the classical contribution 
in this case is smaller than the quantum contribution by a factor of 
$10^{-4}$. This decrease of resistivity would be large enough to observe 
in experiments.

\subsection{Effect of internal field on coherence}
\label{SECmagneticfield}

In ferromagnets the total magnetic field is given by $B=H+4\pi M$, 
where $H$ is the external field and $4\pi M$ is an internal magnetic 
field due to the magnetization. 
In the case of clean Fe wire ($\ell\sim 1.4\mu$m) 
the internal field $4\pi M$ 
is estimated from the of the behavior of the magnetoresistance due to 
the Lorentz motion (at external magnetic field of about 0.1T) as 
$4\pi M\simeq 2.2$T\cite{Kent98}. This large magnetic field can 
destroy the coherence and thus weak localization in the case of film 
or bulk sample\cite{Bergmann84,Lee85}. 
In contrast in mesoscopic wires the effect can be 
neglected if the wire is narrow enough. In fact, in the case of 
magnetization along the wire direction, the magnetic flux penetrating 
though the wire is $\Phi=4\pi M L_{\perp}^2$. The modification of the 
interference of the electron becomes important if this flux is comparable 
to the unit flux $\Phi_{0}\equiv h/2e =2.1\times 10^{-15}$[Tm$^2$]. 
Thus if $L_{\perp}\lesssim 300$\AA, the magnetic flux contained in any 
electron path is too small to affect the coherence. 

The effect of Lorentz motion dut to this internal field
is neglected in disordered case since 
the parameter which determines the effect, $\omega_{c}\tau$ 
($\omega_{c}\equiv eB/m$), is small (for $B=2.2T$ and $\ell=30$\AA, 
$\omega_{c}\tau \simeq 10^{-3}$). 

\section{Conductance fluctuation due the wall motion}
\label{SECCF}

So far we have studied the resistivity due to a static wall.
In this section we will discuss the change of the electron transport due 
to the motion of the wall.
Here we will discuss the conductance of the sample, 
$G\equiv \sigma \times(A_{\perp}/L)$, $A_{\perp}$ being the crosssectional 
area of the wire.
The wall motion can affect the conductance in several different manner, but 
the most significant effect will be that due to the quantum interference.
It has been shown in disordered metal that the fluctuation of the conductance 
can arise at low temperature as a consequence of quantum interference and that 
such fluctuation is of universal order of 
$e^{2}/h= 3.9\times 10^{-5}\Omega^{-1}$ independent of sample size 
and dimensions\cite{LSF85}. 
It has been shown further that if a single impurity atom changes its position 
in such cases, the interference pattern can change and as a result the 
conductance of the entire system can change by 
$O(e^{2}/h)$\cite{Feng86}.
Thus we may expect similar effect of conductance change to arise due to the 
motion of domain walls.
Within the argument of classical resistivity, no change of Boltzmann 
resistivity is expected since it is determined by the reflection 
probability by the wall, which does not depend on the wall position.
If we take into account the quantum interference among the electron, however, 
the motion can affect the conductance change by changing the interference 
pattern.

In the experimental situations, the conductance change of a classical origin 
would also be possible, since the wall motion means the change of the total 
magnetization, and this, according to the argument of the anisotropic 
magnetoresistance\cite{McGuire75}, can lead to the resistivity change.
In this case the expected change for the motion of the wall over a distance 
of $r(\ll L)$ is $\delta \rho \simeq \Delta\rho_{\rm ani}\times (r/L)$, 
$\Delta\rho_{\rm ani}$ being the magnitude of anisotropic 
magnetoresistance; 
$\Delta\rho_{\rm ani}\equiv {\rm Max}(\rho(H))-{\rm Min}(\rho(H))$.
In fact this is what is claimed by Hong and Giordano\cite{Giordano94}
to be the origin of the observed jump in resistivity of Ni wire.
Compared to this classical conductance change, that of a quantum origin 
is much more sensitive to a small motion of a wall. 
Actually it turns out that a displacement of a wall over $100$\AA\ can give 
rise to a conductance change sufficiently large to be observed.

If we write the conductance of the sample with a domain wall at $z=r$ as 
$G(r)$, we are interested in the difference $\delta G\equiv G(r)-G(0)$.
Within the formulation based on Kubo formula, however, 
we cannot directly calculate the conductance $G$ or its change $\delta G$ 
of a fixed sample, since we do not know the 
particular configuration of impurities of the sample. 
We can only calculate the conductance averaged over the impurity configurations, 
in other words, over many samples (this is not so of course in numerical 
calculations). 
Thus we will estimate the mean square of difference of the conductance
\begin{equation}
	<(\delta G)^{2}>_{\rm imp}=
	 2[ <G(0)^{2}>_{\rm imp} -<G(r)G(0)>_{\rm imp} ],
	\label{delGdef}
\end{equation}
where $<>_{\rm imp}$ denotes the average over impurity configurations.
The quantity $<G(r)G(0)>_{\rm imp}$ is expressed as 
\begin{equation}
<G(r)G(0)>_{\rm imp}= \left(\frac{A_{\perp}}{L}\right)^{2}
\lim_{\omega,\omega'\rightarrow0} \frac{1}{\omega\omega'}
 [  F(r,i\omega_{\ell}=\omega+i0,i\omega_{\ell'}=\omega'+i0)
    -F(r,0+i0,0+i0) ],
	\label{delG1}
\end{equation}
where 
\begin{equation}
F(r,i\omega_{\ell},i\omega_{\ell'})
\equiv \frac{\hbar^{2}}{V^{2}} 
<[J(i\omega_{\ell}) J(-i\omega_{\ell})]_{z=r} 
  [J(i\omega_{\ell'}) J(-i\omega_{\ell'})]_{z=0} >.
	\label{Fdef}
\end{equation}
This correlation function is represented diagramatically by two electron 
loops connected through the interaction with the domain wall.
Diagrams with two loops which are not connected by the domain wall line does 
not contribute to $<(\delta G)^{2}>_{\rm imp}$
because this quantity is a difference of the cases with $z=r$ and $z=0$.
The simplest contribution to $<G(r)G(0)>_{\rm imp}$ at the 
lowest, i.e., fourth, order in domain walls is 
that shown in Fig. \ref{FIGCF4}(a) and (a'). 
The domain wall is at $z=r$ in the outer loop and is at $z=0$ in the 
inner loop.
Four Cooperons are included in these processes. 
To calculate the contribution (a), let us first look into the part shown in 
Fig. \ref{FIGdelG}.
The domain wall line here is associated with a factor of $e^{-ipr}$, since 
one of the electron line here interacts with the wall at $z=r$ and the other 
with $z=0$. 
Important contribution comes from the two cases of $n'>0, n<0$ and 
$n'<0, n>0$, in which case the amplitude $\delta\Gamma_{\rm w}$ is written as 
\begin{eqnarray}
\delta\Gamma_{\rm w}(\qv,n'-n) &=& \left(\frac{\hbar^{2}}{2m}\right)^{2}
   (\Gamma_{0}(\qv,n'-n))^{2} \sum_{\kv,p,\sigma}
   \left(k_{z}+\frac{p}{2}\right)
   \left(-k_{z}+q_{z}-\frac{p}{2}\right) |A_{p}|^{2} e^{-ipr}
   \nonumber\\
  &&
\times
   G_{\kv,n,\sigma} G_{\kv+p,n,-\sigma}
   G_{-\kv+\qv,n',\sigma} G_{-\kv+\qv-p,n',-\sigma},
	\label{delG2} 
\end{eqnarray}
where Cooperons $\Gamma_{0}$ is defined in eq.(\ref{cooperon1}).
This is evaluated in the case of $q,p \ll k_{F}$ and the result neglecting 
quantities of $O(\Delta/\epsilon_{F})^{2}$ is
\begin{equation}
 \delta\Gamma _{\rm w} \simeq -\left(\frac{\hbar^{2}}{2m}\right)^{2} 
 \frac{\pi N(0)k_{F}^{2}\tau }{3\hbar\Delta^{2}}\Gamma_{0}^{2} 
 \sum_{p}|A_{p}|^{2}e^{-ipr}.
 	\label{delG3}
\end{equation}
The $p$-summation here results in
\begin{eqnarray}
 	\sum|A_{p}|^{2}e^{-ipr}=\frac{2n_{\rm w}}{\lambda} 
 	\frac{r/\lambda}{\sinh(r/\lambda)} \nonumber\\
 	= \frac{2n_{\rm w}}{\lambda} W\left({r}/{\lambda}\right),
 	\label{sump}
\end{eqnarray}
where 
\begin{equation}
W\left(x\right)\equiv \frac{x}{\sinh x}.
\end{equation}
We thus obtain
 \begin{equation}
 	\delta\Gamma_{\rm w}(\qv,n'-n)\simeq -2\kappa_{\rm w}
 	\frac{n_{\rm i}v^{2}}{\tau^{2}} 
 	(\tilGam(q,n'-n))^{2}W\left({r}/{\lambda}\right),
 	\label{delG4}
 \end{equation}
where 
\begin{equation}
\tilGam(q,n) \equiv \frac{\tau}{n_{\rm i}v^{2}} \Gamma_{0}(q,n) 
=\frac{1}{D\qv^{2}+\omega_{n}},
	\label{tilgam}
\end{equation}
and $\kappa_{\rm w}$ is the dephasing time due to the wall
defined in eq. (\ref{kappaw}). (we suppress here the inelastic lifetime.)
By use of this $\delta \Gamma$, the process of Fig. \ref{FIGCF4}(a) is written 
as ($4$ in the suffix denotes four Cooperons)
\begin{eqnarray}
	F_{4{\rm a}}(\omega_{\ell},\omega_{\ell'}) &=& 
	\left(\frac{e\hbar}{m}\right)^{4} \frac{1}{V^{2}}  
	\sum_{\kv,\kv',\qv,\sigma} \frac{1}{\beta^{2}} \sum_{n,n'}
	k_{z}(-k+q)_{z}k'_{z}(-k'+q)_{z}
	\nonumber\\
&&
\times 	G_{\kv,n+\ell,\sigma} G_{\kv,n,\sigma} 
	G_{-\kv+\qv,n'+\ell',\sigma} G_{-\kv+\qv,n',\sigma}
        G_{\kv',n+\ell,-\sigma} G_{\kv',n,-\sigma} 
	G_{-\kv'+\qv,n'+\ell',-\sigma} G_{-\kv'+\qv,n',-\sigma}
	\nonumber\\
&&
\times  
\delta\Gamma_{\rm w}(q,|n+\ell-n'|) \delta\Gamma_{\rm w}(q,|n-(n'+\ell')|)
        \nonumber\\ 
  &\simeq& F_{0}\frac{W(r/\lambda)^{2}}{\tau^{4}}  
    \frac{1}{\beta^{2}}\sum'_{n,n'} 
      (\tilGam(q,|n+l-n'|))^{2} (\tilGam(q,|n-(n'+\ell')|))^{2} 
        ,
	\label{F2ADEF}
\end{eqnarray}
where
\begin{equation}
	F_{0}\equiv \left[\left(\frac{e\hbar}{m}\right)^{2}
	 \frac{4k_{F}^{2}}{3V}\kappa_{\rm w}\tau^{2}\right]^{2}.
	\label{F0def}
\end{equation}
The summation over the Matsubara frequencies, $\sum'_{n,n'}$, is restricted to 
 the following three cases where the Cooperons has a singular contributions 
($\ell$ and $\ell'$ are positive);
\begin{equation}
 \begin{array}{ccc}
   {\rm I}  :  & n+\ell,n>0,         & n'+\ell',n'<0 \\	
   {\rm II} :  & n+\ell,n<0,         & n'+\ell',n'>0 \\	
   {\rm III}:  & n+\ell,n'+\ell'>0,  & n,n'<0 .
  \end{array}	
\end{equation}
Adding the contributions from these three cases and taking the terms linear 
in both $\omega_{\ell}$ and $\omega_{\ell'}$, which are relevant to the 
conductance change, we obtain (see Appendix \ref{APPCFnsum})
\begin{equation}
\frac{F_{4{\rm a}}} {\omega_{\ell}\omega_{\ell'}} \simeq 
\frac{F_{0}W^{2}}{\tau^{4}}   [ 12J_{6}+J_{6}' ],	
	\label{F2aresult}
\end{equation}
where
\begin{equation}
	J_{m} \equiv \frac{1}{\beta^{2}} \sum_{n,n'=0}^{\infty} 
	\frac{1}{(D\qv^{2}+\omega_{n}+\omega_{n'})^{m}},
	\label{Jmdef}
\end{equation}
and
\begin{eqnarray}
	J_{6}' &\equiv & \frac{1}{\beta^{2}} \sum_{n,n'=0}^{\infty} 
	\left[ -8
	\frac{1}{(D\qv^{2}+\omega_{n}+\omega_{n'})^{3}
         (D\qv^{2}-(\omega_{n}+\omega_{n'}))^{3}} \right.\nonumber\\
     &&   
\left.	+6\left\{ 
\frac{1}{(D\qv^{2}+\omega_{n}+\omega_{n'})^{2}
         (D\qv^{2}-(\omega_{n}+\omega_{n'}))^{4}} \right.\right. 
         \nonumber\\
&& 
\left.\left.
+\frac{1}{(D\qv^{2}+\omega_{n}+\omega_{n'})^{4}
         (D\qv^{2}-(\omega_{n}+\omega_{n'}))^{2}}  \right\} \right]
         .
	\label{Jmpdef}
\end{eqnarray}
It can be shown by similar calculation that the process Fig. \ref{FIGCF4}(a')
leads to a different factor in front of $J_{6}$;
\begin{equation}
 \frac{ F_{4{\rm a'}} } { \omega_{\ell}\omega_{\ell'} } \simeq 
 \frac{ F_{0}W^{2} } { \tau^{4} }   8J_{6}.	
	\label{F2apresult}
\end{equation}
Hence the contributions from the four Cooperons, 
$F_{4}\equiv F_{4{\rm a}} +F_{4{\rm a'}}$, is obtained as
\begin{equation}
\frac{F_{4}}{\omega_{\ell}\omega_{\ell'}} \simeq 
\frac{F_{0}W^{2}}{\tau^{4}}   [ 20J_{6}+J_{6}' ].	
	\label{F2result}
\end{equation}

Processes with five Cooperons are shown in Fig. \ref{FIGCF5}. 
Although they contains two more Cooperons, the contributions turns out to be 
the same order as four Cooperons processes. 
The result of these processes are (Appendix \ref{APPCF56})
\begin{equation}
\frac{F_{5}}{\omega_{\ell}\omega_{\ell'}} =-720 F_{0} 
 W\left({r}/{\lambda}\right)^{2} \frac{Dq_{z}^{2}}{\tau^{4}} 
 J_{7}.
	\label{F5result}
\end{equation}
Similarly contributions from six Cooperons shown in Fig. \ref{FIGCF6} are 
calculated as (Appendix \ref{APPCF56})
\begin{equation}
\frac{F_{6}}{\omega_{\ell}\omega_{\ell'}} = 3024 F_{0} 
 W\left({r}/{\lambda}\right)^{2} \frac{(Dq_{z}^{2})^{2}}{\tau^{4}} 
 J_{8}.
	\label{F6result}
\end{equation}

From these results the total correlation function $F\equiv F_{4}+F_{5}+F_{6}$
is given as
\begin{equation}
\frac{F(i\omega_{\ell},i\omega_{\ell'})}{\omega_{\ell}\omega_{\ell'}}
= \frac{F_{0} W^{2}}{\tau^{4}} \left[ 20J_{6}+J_{6}'-720Dq_{z}^{2} J_{7} 
+3024(Dq_{z}^{2})^{2}J_{8} \right].
	\label{Fresult}
\end{equation}
By use of contour integration $J_{m}$'s are calculated as (Appendix\ref{APPJn})
\begin{equation}
J_{m}=\frac{1}{(2\pi)^{2}(m-1)(m-2)}\frac{1}{(Dq^{2})^{m-2}},
	\label{Jm}
\end{equation}
and similarly $J_{6}'=1/[(2\pi)^{2}(Dq^{2})^{4}]$.
Thus 
\begin{equation}
\frac{F(i\omega_{\ell},i\omega_{\ell'})}{\omega_{\ell}\omega_{\ell'}}
= \frac{F_{0}W^{2}}{(2\pi)^{2}\tau^{4}} \left[ 1+1-8+\frac{72}{5} \right]
\sum_{q} \frac{1}{(Dq^{2})^{4}}.
	\label{Fresult2}
\end{equation}
The summation over $q$ is carried out in one-dimension similarly to eq. 
(\ref{qsumcooperon}).
The result is (recovering $\tau_{\varphi}$ and $\tau_{\rm w}$ in 
Cooperons)
\begin{eqnarray}
\frac{1}{\tau^{4}} \sum_{q} 
\frac{1}{(Dq^{2}+1/\tau_{\varphi}+1/\tau_{\rm w})^{4}} &\simeq& 
 \frac{3^{4} L}{\pi \ell}\int_{0}^{\infty} 
 \frac{dx}{(x^2+3\kappa)^{4}} \nonumber\\
 & =&
\frac{5\sqrt{3}}{32}\frac{L}{\ell}\frac{1}{\kappa^{7/2}},
\label{qsumCF}
\end{eqnarray}
where $\kappa\equiv \kappa_{\varphi}+\kappa_{\rm w}$.
The conductance change is obtained by used of this and eqs. 
(\ref{delGdef})(\ref{delG1})(\ref{Fresult2}) as  
\begin{equation}
\delta G \simeq
\frac{e^{2}}{\hbar} \kappa_{\rm w}
         \sqrt{1-[W\left({r}/{\lambda}\right)]^{2}} 
        \sqrt{\frac{7}{2\sqrt{3}\pi^2}} 
        \sqrt{ \left(\frac{\ell}{L}\right)^3 \frac{1}{\kappa^{7/2}} }.
\end{equation}




\section{Summary}
\label{SECsummary}
We have studied the effect of the domain wall on electronic transport 
properties in wire of ferromagnetic metals. 
We considered the case of 3$d$ transition metals, and took into 
account the scattering by impurities as well.
We have first calculated the conductivity within the classical 
(Boltzmann) transport theory by use of linear response theory. 
This contribution turns out to be negligiblly small in bulk magnets, 
but it can be significant in ballistic nanocontacts, as indicated in 
recent experiments.
Second we discussed the quantum correction of the conductivity 
due to the wall in the disordered case. 
This contribution is due to the dephasing effect caused by the wall 
and thus gives negative contribution to the resistivity. 
this effect grows at lower temperature and 
can win over the classical contribution, in 
particular in wire of diameter $L_{\perp}\lesssim \ell_{\varphi}$, 
$\ell_{\varphi}$ being the inelastic diffusion length. 

So far the studies of transport in magnetic metals has been carried 
out mostly in the system of low resistivity and high temperature, 
partially because of the possible application to devices. 
Our main message here is that in disordered system magnetism can affect 
the transport properties in a novel way by changing the quantum 
coherence among electrons at low temperature. 
At present there are no experiment which clearly indicate this effect, 
but it will be observed in the near future, for instance, in narrower wire 
with a long inelastic diffusion length. 

\section*{Acknowledgements}
The author is indebted to H. Fukuyama for collaboration on the preceeding 
letter, on which this paper is based.
The author thanks G. Bauer, A. Brataas and N. Garcia for collaborations and 
discussions and K. Kuboki, H. Kohno, Y. Otani, K. Takanashi 
for valuable discussions. 
This work is partially supported by a Grand-in-Aid for Scientific 
Research on the Priority Area ``Nanoscale Magnetism and Transport''  
(No.10130219) and ``Spin Controlled Semiconductor 
Nanostructures'' (No.10138211) from the Ministry of Education, Science, 
Sports and Culture.
The author also thanks The Murata Science Foundation and Alexander von Humboldt 
Stiftung for finantial support.
%
\appendix
\section{Calculation of classical conductivity by Kubo formula}
\label{APPQsum}
Here let us give the details of derivation of eq. (\ref{QSUMRESULT}).
Firstly by use of 
\begin{equation}
{k_{z}}(G_{\kv,n,\sigma})^{2}=
      \frac{m}{\hbar^{2}} \frac{\partial}{\partial k_{z}}G_{\kv,n,\sigma},
	\label{partialG}
\end{equation}
and partial integration with respect to $k_{z}$, 
the self-energy term $Q_{2}$ is written as
\begin{eqnarray}
 Q_{2}&=& 
 -\frac{B}{4}\frac{1}{\beta}\sum_{n}\frac{1}{V} 
 \sum_{\kv q \sigma} k_{z}|A_{q}|^{2}\partial_{k_{z}} 
     (G_{\kv,n,\sigma}G_{\kv,n+\ell,\sigma})  \nonumber\\
     &=& \frac{B}{4}\frac{1}{\beta}\sum_{n}\frac{1}{V} 
 \sum_{\kv q \sigma} |A_{q}|^{2}
     G_{\kv,n,\sigma}G_{\kv,n+\ell,\sigma}  ,
     \label{Q21}
\end{eqnarray}
where $B\equiv ((e\hbar)^2/2m^{2})$.
The other self-energy contribution, $Q_{4}$, can be rewritten by use 
of the 
identity
\begin{equation}
G_{\kv,n,\sigma}G_{\kv,n+\ell,\sigma}=
	\frac{1}{i\left(\omega_{\ell}+\Delta_{n\ell}\right)}
	(G_{\kv,n,\sigma}-G_{\kv,n+\ell,\sigma}),
	\label{identity1}
\end{equation}
where 
\begin{equation}
\Delta_{n\ell}\equiv \frac{\hbar}{2\tau}({\rm sgn}(n+\ell)-{\rm 
sgn}(n)),
	\label{Dnldef}
\end{equation}
as
\begin{eqnarray}
	Q_{4}&=& -\frac{B\hbar^{4}}{2m^{2}}  \frac{1}{\beta}\sum_{n}
\frac{1}{V}\sum_{\kv q\sigma }
k_{z}^{2}\left(k_{z}-\frac{q}{2}\right)^{2} |A_{q}|^{2}
\frac{1}{i\left(\omega_{\ell}+\Delta_{n\ell}\right)}  \nonumber\\
&&\times
[(G_{\kv-\frac{q}{2},n,\sigma}-G_{\kv-\frac{q}{2},n+\ell,\sigma})
 (G_{\kv-\frac{q}{2},n,\sigma}G_{\kv+\frac{q}{2},n,-\sigma}  
 +
G_{\kv-\frac{q}{2},n+\ell,\sigma}G_{\kv+\frac{q}{2},n+\ell,-\sigma}  
) ]
	\nonumber\\
&=& 
-\frac{B\hbar^{4}}{2m^{2}}  \frac{1}{\beta}\sum_{n}
\frac{1}{V}\sum_{\kv q\sigma }
k_{z}^{2}\left(k_{z}-\frac{q}{2}\right)^{2} |A_{q}|^{2}
\frac{1}{i\left(\omega_{\ell}+\Delta_{n\ell}\right)}  
\nonumber\\
&&\times 
[-G_{\kv-\frac{q}{2},n,\sigma}G_{\kv-\frac{q}{2},n+\ell,\sigma}
(G_{\kv+\frac{q}{2},n,-\sigma}-G_{\kv+\frac{q}{2},n+\ell,-\sigma}) 
\nonumber\\
&&
+ (G_{\kv-\frac{q}{2},n,\sigma})^{2}G_{\kv+\frac{q}{2},n,-\sigma}
- 
(G_{\kv-\frac{q}{2},n+\ell,\sigma})^{2}G_{\kv+\frac{q}{2},n+\ell,-\sigma}]
\nonumber\\
&\equiv &Q_{4a}+Q_{4b} ,	\label{Q41}
\end{eqnarray}
where
\begin{eqnarray}
	Q_{4a}&\equiv& \frac{B\hbar^{4}}{2m^{2}}\frac{1}{\beta}\sum_{n}
\frac{1}{V}\sum_{\kv q\sigma }
k_{z}^{2}\left(k_{z}-\frac{q}{2}\right)^{2} |A_{q}|^{2}
G_{\kv-\frac{q}{2},n,\sigma}
G_{\kv-\frac{q}{2},n+\ell,\sigma}
G_{\kv+\frac{q}{2},n,-\sigma}
G_{\kv+\frac{q}{2},n+\ell,-\sigma}  ,\nonumber\\
          Q_{4b}&\equiv& -\frac{B\hbar^{4}}{2m^{2}}\frac{1}{\beta}\sum_{n}
\frac{1}{V}\sum_{\kv q\sigma }
k_{z}^{2}\left(k_{z}-\frac{q}{2}\right)^{2} |A_{q}|^{2}
\frac{1}{i\left(\omega_{\ell}+\Delta_{n\ell}\right)}
\nonumber\\
&&
\times \left[ 
(G_{\kv-\frac{q}{2},n,\sigma})^{2}G_{\kv+\frac{q}{2},n,-\sigma}  
-(G_{\kv-\frac{q}{2},n+\ell,\sigma})^{2}G_{\kv+\frac{q}{2},n+\ell,-\sigma} 
\right].
	\label{Q42}
\end{eqnarray}
The sum of the terms $Q_{4a}+Q_{5}$ can then be simplified as
\begin{equation}
	Q_{4a}+Q_{5}=
	\frac{B \hbar^{4}}{4m^{2}}\frac{1}{\beta}\sum_{n}
\frac{1}{V}\sum_{\kv q\sigma }
(k_{z}q)^{2} |A_{q}|^{2}
G_{\kv-\frac{q}{2},n,\sigma}
G_{\kv-\frac{q}{2},n+\ell,\sigma}
G_{\kv+\frac{q}{2},n,-\sigma}
G_{\kv+\frac{q}{2},n+\ell,-\sigma} .
	\label{Q4a5}
\end{equation}
This expression can be rewritten further by use of the identity
\begin{equation}
\left(\frac{\hbar^{2} k_{z}q}{m}+2\sigma \Delta \right) 
G_{\kv+\frac{q}{2},n,-\sigma}
                                    G_{\kv-\frac{q}{2},n,\sigma}
=	G_{\kv+\frac{q}{2},n,-\sigma}-G_{\kv-\frac{q}{2},n,\sigma},
	\label{identity2}
\end{equation}
and the expression (\ref{Q21}) for $Q_{2}$ as
\begin{eqnarray}
\lefteqn{	Q_{4a}+Q_{5}= \frac{B}{4m^{2}}
	\frac{1}{\beta}\sum_{n}\frac{1}{V}\sum_{\kv q\sigma }|A_{q}|^{2}
 [
 (2m\Delta)^{2} 
G_{\kv-\frac{q}{2},n,\sigma}
G_{\kv-\frac{q}{2},n+\ell,\sigma}
G_{\kv+\frac{q}{2},n,-\sigma}
G_{\kv+\frac{q}{2},n+\ell,-\sigma} 
}
\nonumber\\
&&
+2m^{2}\sigma\Delta  
 \{ G_{\kv-\frac{q}{2},n,\sigma}G_{\kv+\frac{q}{2},n,-\sigma}
(G_{\kv-\frac{q}{2},n+\ell,\sigma}-G_{\kv+\frac{q}{2},n+\ell,-\sigma})
 \nonumber\\
&&
 +G_{\kv-\frac{q}{2},n+\ell,\sigma}G_{\kv+\frac{q}{2},n+\ell,-\sigma}
     (G_{\kv-\frac{q}{2},n,\sigma}-G_{\kv+\frac{q}{2},n,-\sigma})    
\}    
\nonumber\\
&&
+m^{2} (G_{\kv-\frac{q}{2},n,\sigma}-G_{\kv+\frac{q}{2},n,-\sigma})
(G_{\kv-\frac{q}{2},n+\ell,\sigma}-G_{\kv+\frac{q}{2},n+\ell,-\sigma})
 ]
\nonumber\\
&=& 	Q_{\rm c}+Q_{1}+2Q_{2}+Q_{45}',
	\label{Q4a51}
\end{eqnarray}
where 
\begin{equation}
Q_{\rm c}\equiv B\Delta^{2}
\frac{1}{\beta}\sum_{n}\frac{1}{V}\sum_{\kv q\sigma } |A_{q}|^{2} 
G_{\kv-\frac{q}{2},n,\sigma}
G_{\kv-\frac{q}{2},n+\ell,\sigma}
G_{\kv+\frac{q}{2},n,-\sigma}
G_{\kv+\frac{q}{2},n+\ell,-\sigma},
	\label{QMdef}
\end{equation}
is the term which survives in the final expression, 
(\ref{QSUMRESULT}), and 
\begin{equation}
	Q_{45}'\equiv 	B\frac{1}{\beta}\sum_{n}
\frac{1}{V}\sum_{\kv q\sigma }
\sigma\Delta |A_{q}|^{2}
[ G_{\kv-\frac{q}{2},n,\sigma} G_{\kv+\frac{q}{2},n,-\sigma}
                              G_{\kv-\frac{q}{2},n+\ell,\sigma}
    +
  G_{\kv-\frac{q}{2},n+\ell,\sigma} G_{\kv+\frac{q}{2},n+\ell,-\sigma}
                              G_{\kv-\frac{q}{2},n,\sigma}  ].
	\label{Q45prime}
\end{equation}
Similarly by use of identities (\ref{identity1}) and (\ref{identity2}), 
$Q_{4b}$ of (\ref{Q42}) is rewritten as
\begin{eqnarray}
Q_{4b} &=& 
-\frac{B \hbar^{2}}{2m}\frac{1}{\beta}\sum_{n}
\frac{1}{V}\sum_{\kv q\sigma }
k_{z}^{2}\left(k_{z}-\frac{q}{2}\right) |A_{q}|^{2}
\frac{1}{i\left(\omega_{\ell}+\Delta_{n\ell}\right)}
\nonumber\\
&&
\times \left[ 
( 
\partial_{k_{z}}G_{\kv-\frac{q}{2},n,\sigma})G_{\kv+\frac{q}{2},n,-\sigma} 
-(\partial_{k_{z}}G_{\kv-\frac{q}{2},n+\ell,\sigma})
           G_{\kv+\frac{q}{2},n+\ell,-\sigma} 
  \right] \nonumber\\
&=&
-\frac{B\hbar^{2}}{2m}\frac{1}{\beta}\sum_{n}
\frac{1}{V}\sum_{\kv q\sigma }
k_{z}^{2} |A_{q}|^{2}
\frac{1}{i\left(\omega_{\ell}+\Delta_{n\ell}\right)}
\nonumber\\
&&
\times \frac{1}{2} \left[ \left(k_{z}-\frac{q}{2}\right)\left[ 
( 
\partial_{k_{z}}G_{\kv-\frac{q}{2},n,\sigma})G_{\kv+\frac{q}{2},n,-\sigma} 
-(\partial_{k_{z}}G_{\kv-\frac{q}{2},n+\ell,\sigma})
           G_{\kv+\frac{q}{2},n+\ell,-\sigma} 
  \right]  \right. \nonumber\\
&&
\left. 
+ \left(k_{z}+\frac{q}{2}\right)\left[  
G_{\kv-\frac{q}{2},n,\sigma} \partial_{k_{z}}G_{\kv+\frac{q}{2},n,-\sigma} 
-G_{\kv-\frac{q}{2},n+\ell,\sigma}
           \partial_{k_{z}}G_{\kv+\frac{q}{2},n+\ell,-\sigma} 
  \right] \right] \nonumber\\
&=&  
 -\frac{B\hbar^{2}}{4m}\frac{1}{\beta}\sum_{n}
\frac{1}{V}\sum_{\kv q\sigma }
k_{z}^{3}|A_{q}|^{2}
\frac{1}{i\left(\omega_{\ell}+\Delta_{n\ell}\right)}
\nonumber\\
&&
\times  \partial_{k_{z}} \left[ 
G_{\kv-\frac{q}{2},n,\sigma}G_{\kv+\frac{q}{2},n,-\sigma} 
 - 
G_{\kv-\frac{q}{2},n+\ell,\sigma}G_{\kv+\frac{q}{2},n+\ell,-\sigma} 
  \right] 
 \nonumber\\
&+&  
   \frac{B\hbar^{4}}{4m^{2}}\frac{1}{\beta}\sum_{n}
\frac{1}{V}\sum_{\kv q\sigma }
k_{z}^{2}q\left(k_{z}-\frac{q}{2}\right) |A_{q}|^{2}
\frac{1}{i\left(\omega_{\ell}+\Delta_{n\ell}\right)}
\nonumber\\
&&
\times \left[ 
(G_{\kv-\frac{q}{2},n,\sigma})^{2}G_{\kv+\frac{q}{2},n,-\sigma} 
 -(G_{\kv-\frac{q}{2},n+\ell,\sigma})^{2}G_{\kv+\frac{q}{2},n+\ell,-\sigma} 
\right] 
\nonumber\\
&\equiv&
Q_{4b}'+Q_{4b}''.
\label{Q4b1}
\end{eqnarray}
The first term ($\equiv Q_{4b}'$) turns out by use of partial integration to be
\begin{eqnarray}
Q_{4b}'&=& 
\frac{3B\hbar^{2}}{4m}\frac{1}{\beta}\sum_{n}
\frac{1}{V}\sum_{\kv q\sigma }
k_{z}^{2}|A_{q}|^{2}
\frac{1}{i\left(\omega_{\ell}+\Delta_{n\ell}\right)}
\left[ 
G_{\kv-\frac{q}{2},n,\sigma}G_{\kv+\frac{q}{2},n,-\sigma} 
 -G_{\kv-\frac{q}{2},n+\ell,\sigma}
           G_{\kv+\frac{q}{2},n+\ell,-\sigma} 
  \right] 
 \nonumber\\
&=& 
-\frac{3}{4}Q_{3a},
\end{eqnarray}
where $Q_{3a}$ is the term which appears in the expression of 
$Q_{3}$, 
\begin{eqnarray}
	Q_{3a}&\equiv & 
-\frac{B \hbar^{2}}{m}\frac{1}{\beta}\sum_{n}
\frac{1}{V}\sum_{\kv q\sigma}
k_{z}^{2}|A_{q}|^{2}
[
G_{\kv-\frac{q}{2},n,\sigma}
G_{\kv-\frac{q}{2},n+\ell,\sigma}
G_{\kv+\frac{q}{2},n,-\sigma}
+
G_{\kv-\frac{q}{2},n+\ell,\sigma}
G_{\kv+\frac{q}{2},n,-\sigma}
G_{\kv+\frac{q}{2},n+\ell,-\sigma}
].
\nonumber\\
&&
	\label{Q3adef}
\end{eqnarray}
The second term in (\ref{Q4b1}) is rewritten by use of 
(\ref{identity2}) and partial integration as
$Q_{4b}''=Q_{2}-(1/4)Q_{3a}+Q_{\rm c}'$ where 
\begin{equation}
Q_{\rm c}'\equiv -\frac{B}{2} \frac{1}{\beta}\sum_{n}
\frac{1}{V}\sum_{\kv q\sigma}
\sigma\Delta k_{z}|A_{q}|^{2}
\frac{1}{i\left(\omega_{\ell}+\Delta_{n\ell}\right)}
\left[
G_{\kv+\frac{q}{2},n,-\sigma}
\partial_{k_{z}} G_{\kv-\frac{q}{2},n,\sigma}
\right. 
-
\left. G_{\kv+\frac{q}{2},n+\ell,-\sigma}
\partial_{k_{z}} G_{\kv-\frac{q}{2},n+\ell,\sigma}
\right].
\end{equation}
Thus we obtain 
\begin{equation}
	Q_{4b}= Q_{4b}'+Q_{4b}''=-Q_{3a}+Q_{2}+Q_{\rm c}'.
	\label{Q4b2}
\end{equation}
Similarly $Q_{3}$ is written as
\begin{eqnarray}
Q_{3}&=&Q_{3a}+\frac{B\hbar^{2}}{2m}\frac{1}{\beta}\sum_{n}
\frac{1}{V}\sum_{\kv q\sigma}
k_{z}q |A_{q}|^{2}
[
G_{\kv-\frac{q}{2},n,\sigma}
G_{\kv-\frac{q}{2},n+\ell,\sigma}
G_{\kv+\frac{q}{2},n,-\sigma}
+
G_{\kv-\frac{q}{2},n,\sigma}
G_{\kv-\frac{q}{2},n+\ell,\sigma}
G_{\kv+\frac{q}{2},n+\ell,-\sigma}
]
\nonumber\\
&=&
Q_{3a}-4Q_{2}-2Q_{1}-Q_{45}'.
	\label{Q31}
\end{eqnarray}
Adding the three equations (\ref{Q4a51})(\ref{Q4b2}) and 
(\ref{Q31}), we finally obtain the result 
\begin{equation}
	Q_{1}+Q_{2}+Q_{3}+Q_{4}+Q_{5}=Q_{\rm c}+Q_{\rm c}'.
	\label{Qsumresult2}
\end{equation}

The term $Q_{\rm c}'$ is rewritten by use of partial integration over $q$ as 
\begin{eqnarray}
\lefteqn{ Q_{\rm c}'= \frac{B}{2}\frac{1}{\beta}\sum_{n}
\frac{1}{V}\sum_{\kv q\sigma}
\sigma\Delta k_{z}|A_{q}|^{2}
\frac{1}{i\left(\omega_{\ell}+\Delta_{n\ell}\right)}
\partial_{q} \left[
G_{\kv+\frac{q}{2},n,-\sigma}
G_{\kv-\frac{q}{2},n,\sigma}
-
 G_{\kv+\frac{q}{2},n+\ell,-\sigma}
 G_{\kv-\frac{q}{2},n+\ell,\sigma}
\right]
}
\nonumber\\
&&=
-\frac{B}{2} \frac{1}{\beta}\sum_{n}
\frac{1}{V}\sum_{\kv q\sigma}
\sigma\Delta k_{z}(\partial_{q} |A_{q}|^{2})
\frac{1}{i\left(\omega_{\ell}+\Delta_{n\ell}\right)}
\left[
G_{\kv+\frac{q}{2},n,-\sigma}
G_{\kv-\frac{q}{2},n,\sigma}
-
 G_{\kv+\frac{q}{2},n+\ell,-\sigma}
G_{\kv-\frac{q}{2},n+\ell,\sigma}
\right]
.  \nonumber\\
&&
\end{eqnarray}
By use of (\ref{identity1}) and (\ref{identity2}) this becomes
\begin{equation}
Q_{\rm c}'= \frac{B}{2} \frac{1}{\beta}\sum_{n}
\frac{1}{V}\sum_{\kv q\sigma}
(\partial_{q} |A_{q}|^{2})
\frac{\Delta \left( k_{z}+\frac{q}{2} 
\right)}{\Delta+\sigma\frac{(k_{z}+q/2)q}{m}}
G_{\kv,n,\sigma}G_{\kv,n+\ell,\sigma}.
\end{equation}

\section{Conductivity corrections $\delta \sigma$ and $\sigma'$}
\label{APPdeln}
First we calculated the conductivity correction due to the shift of 
the electron density.
The correction to the electron density due to the interaction 
(\ref{Hint}) is given by $\delta n\equiv \delta n_{1}+\delta n_{2}$, 
where
\begin{equation}
	\delta n_{1}=\frac{\hbar^{2}}{4m}\frac{1}{\beta V}\sum_{\kv q n\sigma} 
|A_{q}|^{2} \frac{1}{2}(G_{\kv n\sigma})^{2} ,
\end{equation}
and
\begin{equation}
	\delta n_{2}=\frac{\hbar^{2}}{4m}\frac{1}{\beta V}\sum_{\kv q n\sigma} 
|A_{q}|^{2} \frac{\hbar^{2} k_{z}^{2}}{m}
 (G_{\kv-\frac{q}{2}, n\sigma})^{2} G_{\kv+\frac{q}{2}, n,-\sigma} .
\end{equation}
These contributions are shown diagramatically in Fig. \ref{FIGdeltan}.
By use of (\ref{identity2}) the expression of $\delta n_{2}$ 
is simplified to be
\begin{equation}
\delta n_{2} = -\frac{\hbar^{2}}{4m^{2}} \frac{1}{\beta V}\sum_{\kv q n\sigma} 
|A_{q}|^{2} 
\frac{ \left(k_{z}+\frac{q}{2}\right)^{2} }
{ 2\sigma\Delta+\frac{(k_{z}+q/2)q}{m} }
 (G_{\kv, n\sigma})^{2} .
\end{equation}
The summation over $\omega_{n}$ is carried out as (see Appendix \ref{APPomegasum}) 
\begin{eqnarray}
	\frac{1}{\beta}\sum_{n} (G_{\kv, n\sigma})^{2} &=& -\frac{1}{2\pi i} 
	\int_{-\infty}^{\infty} dz f(z)\frac{i}{\tau} \frac{d}{dz} 
	\frac{1}{(z-\epsilon_{\kv\sigma})^{2}+\left(\frac{\hbar}{2\tau}\right)^{2}}
	\nonumber\\
	&=& -\frac{1}{2\pi \tau} 
	\frac{1}{\epsilon_{\kv\sigma}^{2}+\left(\frac{\hbar}{2\tau}\right)^{2}}.
\end{eqnarray} 
Thus we finally obtain 
\begin{equation}
\delta n= -\frac{\hbar^{3}}{16\pi m\tau }\frac{1}{V}\sum_{\kv q\sigma} 
\frac{|A_{q}|^{2}}
 {\epsilon_{\kv\sigma}^{2}+\left(\frac{\hbar}{2\tau}\right)^{2}}
 \frac{\Delta-\sigma\frac{(k_{z}+q/2)k_{z}}{m}} 
{\Delta+\sigma\frac{(k_{z}+q/2)q}{2m} }.
\end{equation}

Thus the correction of the zeroth order conductivity, 
$\delta\sigma_{\rm c}=e^{2}\tau\delta n /m$ 
is obtained as
\begin{equation}
\delta \sigma_{\rm c}= -\frac{\hbar}{16\pi } \left( 
\frac{e\hbar}{m}\right)^{2} \frac{1}{V}\sum_{\kv q\sigma} 
\frac{|A_{q}|^{2}}
 {\epsilon_{\kv\sigma}^{2}+\left(\frac{\hbar}{2\tau}\right)^{2}}
 \frac{\Delta-\sigma\frac{(k_{z}+q/2)k_{z}}{m}} 
{\Delta+\sigma\frac{(k_{z}+q/2)q}{2m} }.
\label{delsigapp}
\end{equation}
The $\kv$-summation is easily carried out by rewriting it by the 
energy integration to obtain
\begin{equation}
\delta \sigma_{\rm c}= -\frac{\hbar }{8} \left( 
\frac{e\hbar}{m}\right)^{2} \frac{1}{V}\sum_{q\sigma}  N_{\sigma}\tau 
|A_{q}|^{2} \left<
 \frac{\Delta-\sigma\frac{(k_{z}+q/2)k_{z}}{m}} 
{\Delta+\sigma\frac{(k_{z}+q/2)q}{2m} } \right>,
\label{delsigresultapp}
\end{equation}
where bracket denotes the angular average over 
$k_{z}\equiv k_{F\sigma}\cos\theta$.

Similarly the contribution $\sigma_{\rm c}'$, eq. (\ref{sigmap}), is calculated as
\begin{equation}
\sigma_{\rm c}'= \frac{\hbar}{8} \left( 
\frac{e\hbar}{m}\right)^{2} \frac{1}{V}\sum_{q\sigma}  N_{\sigma}\tau 
|A_{q}|^{2}  \left<
 \frac{\Delta\left(\Delta-\sigma\frac{(k_{z}+q/2)^{2}}{m} \right) } 
{ \left[\Delta+\sigma\frac{(k_{z}+q/2)q}{2m} \right]^{2}} \right>.
	\label{sigmapapp}
\end{equation}
From these expressions (\ref{delsigresultapp}) and (\ref{sigmapapp})
it is seen that in the limit of 
$k_{F}\lambda\gg1$ (i.e., $q \ll k_{F}$) and 
$\Delta \gg \hbar^{2} k_{F}/m\lambda$, the sum of two terms vanishes; 
$\sigma_{\rm c}'+\delta\sigma_{\rm c}=0$.

\section{Summation over $\omega_{n}$ in eq. (\protect\ref{QSUMRESULT})}
\label{APPomegasum}
The summation over Matsubara frequency, $\omega_{n}$, in the 
expression of classical contribution to the conductivity, 
(\ref{QSUMRESULT}), can be carried out by use of contour integration 
as follows.
The quantity we consider is 
\begin{equation}
	Q_{\rm c}(i\omega_{\ell})=-B\frac{1}{V}\sum_{\kv q\sigma } 
|A_{q}|^{2} I_{\kv q\sigma }(i\omega_{\ell}),
	\label{Qc1}
\end{equation}
where $B=((e\hbar)^{2}/2m^{2})$ and 
\begin{equation}
I_{\kv q\sigma }(i\omega_{\ell})\equiv 
\frac{1}{\beta}\sum_{n}
G_{\kv-\frac{q}{2},n,\sigma}
G_{\kv-\frac{q}{2},n+\ell,\sigma}
G_{\kv+\frac{q}{2},n,-\sigma}
G_{\kv+\frac{q}{2},n+\ell,-\sigma}.
	\label{Idef}
\end{equation}
This function is written by use of contour integration with respect 
to $z\equiv i\omega_{n}$ as
\begin{equation}
	I_{\kv q\sigma }(i\omega_{\ell})=
-\frac{1}{2\pi i}\oint_{C_{0}} dz f(z)
G_{\kv-\frac{q}{2},\sigma}(z)
G_{\kv-\frac{q}{2},\sigma}(z+i\omega_{\ell})
G_{\kv+\frac{q}{2},-\sigma}(z)
G_{\kv+\frac{q}{2},-\sigma}(z+i\omega_{\ell}),
	\label{I1}
\end{equation}
where $f(z)\equiv 1/(1+e^{\beta z})$ and Green function $G_{\kv,\sigma,n}$ is 
here denoted by $G_{\kv,\sigma}(i\omega_{n})$, and the contour $C_{0}$ goes 
around the imaginary axis in complex $z$-plane (Fig. \ref{FIGcontour}).
Noting that 
$G_{\kv,\sigma}(z)=[z+(1/2\tau){\rm sgn}({\rm Im}[z])-\epsilon_{\kv,\sigma}]^{-1}$
has a cut along ${\rm Im}[z]=0$, the contour can be changed to four 
paths $C_{1}$ parallel to the real axis, namely 
$z\equiv \pm(\omegap+i0)$ and $z\equiv 
-i\omega_{\ell}\pm(\omegap+i0)$,
where $\omegap$ runs from $-\infty$ to $\infty$ (Fig. \ref{FIGcontour}).                                                                                    
We then obtain
\begin{eqnarray}
I_{\kv q\sigma }(i\omega_{\ell})&=&
-\frac{1}{\pi}\int_{-\infty}^{\infty} d\omegap 
\left[ f(\omegap) 
 \frac{1}{(\omegap+i\omega_{\ell}+\frac{i\hbar}{2\tau}-\epsilon_{\kvm,\sigma})}
 \frac{1}{(\omegap+i\omega_{\ell}+\frac{i\hbar}{2\tau}-\epsilon_{\kvp,-\sigma})}
\right.
    \nonumber\\
   && \times {\rm Im} \left(
      \frac{1}{(\omegap+\frac{i\hbar}{2\tau}-\epsilon_{\kvm,\sigma})}
     \frac{1}{(\omegap+\frac{i\hbar}{2\tau}-\epsilon_{\kvp,-\sigma})}
     \right)
     \nonumber\\
  &&  +f(\omegap-i\omega_{\ell}) 
 \frac{1}{(\omegap-i\omega_{\ell}-\frac{i\hbar}{2\tau}-\epsilon_{\kvm,\sigma})}
\frac{1}{(\omegap-i\omega_{\ell}-\frac{i\hbar}{2\tau}-\epsilon_{\kvp,-\sigma})}
 \nonumber\\
   && \left.
     \times {\rm Im} \left(
      \frac{1}{(\omegap+\frac{i\hbar}{2\tau}-\epsilon_{\kvm,\sigma})}
     \frac{1}{(\omegap+\frac{i\hbar}{2\tau}-\epsilon_{\kvp,-\sigma})}
     \right)
     \right]
 . \label{I2}
\end{eqnarray}
The classical conductivity is expressed by taking the imaginary part of the 
correlation function analytically continued to 
$\omega+i0\equiv i\omega_{\ell}$ as
\begin{equation}
	\sigma_{\rm w}=\lim_{\omega\rightarrow 0}{\rm Im} 
	\frac{Q_{\rm  w}(\omega+i0)-Q_{\rm  w}(i0)}{\omega}.
	\label{sigmacdef}
\end{equation}
The imaginary part of $I_{\kv q\sigma }(i\omega_{\ell}=\omega+i0)$ is 
obtained from 
(\ref{I2}) as 
\begin{eqnarray}
{\rm Im}I_{\kv q\sigma }(\omega+i0)&=&
-\frac{1}{\pi}\int_{-\infty}^{\infty} d\omegap 
   {\rm Im} \left(
       \frac{1}{(\omegap+\frac{i\hbar}{2\tau}-\epsilon_{\kvm,\sigma})}
      \frac{1}{(\omegap+\frac{i\hbar}{2\tau}-\epsilon_{\kvp,-\sigma})}
      \right)
      \nonumber\\
&&\times \left[ 
   f(\omegap) {\rm Im} \left( 
\frac{1}{(\omegap+\omega+\frac{i\hbar}{2\tau}-\epsilon_{\kvm,\sigma})}
\frac{1}{(\omegap+\omega+\frac{i\hbar}{2\tau}-\epsilon_{\kvp,-\sigma})}
\right.\right.
 \nonumber\\
&&
- \left.\left.
f(\omegap-\omega)
\frac{1}{(\omegap-\omega+\frac{i\hbar}{2\tau}-\epsilon_{\kvm,\sigma})}
\frac{1}{(\omegap-\omega+\frac{i\hbar}{2\tau}-\epsilon_{\kvp,-\sigma})}
      \right)
      \right]
    \nonumber\\
 &=& 
 -\frac{1}{\pi}\int_{-\infty}^{\infty} d\omegap 
\left[ f(\omegap) -f(\omegap+\omega) \right]
{\rm Im} \left( 
\frac{1}{\omegap+\omega+\frac{i\hbar}{2\tau}-\epsilon_{\kvm,\sigma}}
\frac{1}{\omegap+\omega+\frac{i\hbar}{2\tau}-\epsilon_{\kvp,-\sigma}}
            \right)
\nonumber\\
&&\times
        {\rm Im} \left(
       \frac{1}{\omegap+\frac{i\hbar}{2\tau}-\epsilon_{\kvm,\sigma}}
      \frac{1}{\omegap+\frac{i\hbar}{2\tau}-\epsilon_{\kvp,-\sigma}}
      \right)
 . \label{I3}
\end{eqnarray}
By use of 
\begin{equation}
	f(\omegap)-f(\omegap+\omega)= -\omega\frac{d f(\omegap)}{d\omegap} 
	        +O(\omega^{2})
	     \sim \omega\delta(\omegap)   ,
	\label{delfrel}
\end{equation}
which holds at small $\omega$ and low temperature, we obtain 
\begin{equation}
	{\rm Im}I_{\kv q\sigma }(\omega+i0) \sim
	-\frac{\omega}{\pi}
        \left[  {\rm Im} \left( 
       \frac{1}{\frac{i\hbar}{2\tau}-\epsilon_{\kvm,\sigma}}
      \frac{1}{\frac{i\hbar}{2\tau}-\epsilon_{\kvp,-\sigma}} 
      \right) \right]^{2}.	
      \label{ImI}
\end{equation}
The classical contribution to the conductivity, 
(\ref{sigmacdef}),  is then obtained as 
\begin{equation}
	\sigma_{\rm c}=
-\frac{\Delta^{2}\hbar^{3}}{8\pi\tau^{2}}\left(\frac{e\hbar}{m}\right)^{2}
\frac{1}{V}\sum_{\kv q\sigma }|A_{q}|^{2}
\frac{(\epsilon_{\kv-\frac{q}{2},\sigma}
       +\epsilon_{\kv+\frac{q}{2},-\sigma})^{2}}   
   {\left[(\epsilon_{\kv-\frac{q}{2},\sigma})^{2}
         +\left(\frac{\hbar}{2\tau}\right)^{2}\right]^{2}
    \left[(\epsilon_{\kv+\frac{q}{2},-\sigma})^{2}
         +\left(\frac{\hbar}{2\tau}\right)^{2}\right]^{2}}.
\label{sigmacresult2}
\end{equation}
\section{Derivation of eq. (\protect\ref{SIGMAC}) and 
(\protect\ref{SIGMACFERRO0}) }
\label{APPksum}
The $\kv$-summation in (\ref{sigmacresult}) is carried out as follows.
We neglect quantities of $O((q/k_{F})^{2})$ and 
approximate $\epsilon_{\kv\pm q/2,\mp\sigma}\simeq 
\epsilon_{\kv}\pm[(k_{z}q/2m)+\sigma\Delta]$.
This is because the momentum transfer, $q$, is limited to a small value of 
$q\lesssim \lambda^{-1}$ due to the 
form factor of the wall, $|A_{q}|^{2}\propto [\cosh(\pi q\lambda/2)]^{-2}$, and we are considering the 
case of a thick wall, $k_{F}\lambda \gg 1$.
Then $\sigma_{\rm w}$  is written as
\begin{equation}
       \sigma_{\rm c} \simeq 
	-\frac{\Delta^{2}\hbar^{3}}{8\pi\tau^{2}}\left(\frac{e\hbar}{m}\right)^{2}
	\frac{1}{V}\sum_{q}|A_{q}|^{2}J_{q},
	\label{sigamcJ}
\end{equation}
where
\begin{equation}
	J_{q}\equiv\sum_{{\kv\sigma}} 
	\frac{4(\epsilon_{\kv})^{2}}
	        {   \left[   \left\{
	\epsilon_{\kv}-\left(\frac{\hbar^{2}k_{z}q}{2m}+\sigma\Delta\right)
	    \right\}^{2} +\left(\frac{\hbar}{2\tau}\right)^{2}  \right]^{2}
                      \left[   \left\{
	\epsilon_{\kv}+\left(\frac{\hbar^{2}k_{z}q}{2m}+\sigma\Delta\right)
	    \right\}^{2} +\left(\frac{\hbar}{2\tau}\right)^{2}  \right]^{2}
	    }.
	\label{Jqdef}
\end{equation}
This function is written as
\begin{eqnarray}
J_{q}&\equiv & 2\int_{-1}^{1}\frac{d\cos\theta}{2} 
\int_{-\epsilon_{F}}^{\infty}d\epsilon N(\epsilon)
\frac{4\epsilon^{2}}
	        {   \left[   \left\{
	  \epsilon-\left(\frac{\hbar^{2}kq\cos\theta}{2m}+\Delta\right)
	    \right\}^{2} +\left(\frac{\hbar}{2\tau}\right)^{2}  \right]^{2}
                      \left[   \left\{
	  \epsilon+\left(\frac{\hbar^{2}kq\cos\theta}{2m}+\Delta\right)
	    \right\}^{2} +\left(\frac{\hbar}{2\tau}\right)^{2}  \right]^{2}
	    }
	    \nonumber\\
&\equiv& \int_{-1}^{1}{d\cos\theta}
\int_{-\epsilon_{F}}^{\infty}d\epsilon N(\epsilon) F_{\theta}(\epsilon),
\label{Jq1}
\end{eqnarray}
where $N(\epsilon)\equiv 
(Vm^{3/2}/\pi^{2}\sqrt{2}\hbar^{3})\sqrt{\epsilon+\epsilon_{F}}$ is the 
density of states, and  
$k\equiv\sqrt{2m(\epsilon+\epsilon_{F})}/\hbar$, $\theta$ being the 
angle between $\kv$ and the $z$-axis, and $F_{\theta}(\epsilon)$ is 
defined by the last line.
In the first line we have included the factor of two due to the 
summation over the spin.
The integration over $\epsilon$ can be carried out by use of contour 
integration. 
In doing this we need to be a little bit careful 
since the density of states in three-dimensions 
$N(\epsilon)\propto\sqrt{\epsilon+\epsilon_{F}}$ has a cut on the real 
axis running from $\epsilon=-\epsilon_{F}$ to $\epsilon={\infty}$.
Choosing a closed path $C_{2}$ in the $\epsilon$-plane as in Fig. 
\ref{FIGContour2}, 
the $\epsilon$-integral in (\ref{Jq1}) is written as
\begin{equation}
\int_{-\epsilon_{F}}^{\infty}d\epsilon N(\epsilon) F_{\theta}(\epsilon)
=\frac{1}{2} \oint_{C_{2}} d\epsilon N(\epsilon) F_{\theta}(\epsilon).
	\label{Jq2}
\end{equation}
The contour $C_{2}$ contains four poles at 
$\epsilon=\epsilon^{*}_{\sigma,\pm}\equiv 
\sigma\left(\Delta+\frac{\hbar^{2}k(\epsilon^{*}_{\sigma})q}{2m} 
\cos\theta\right)\pm i\frac{\hbar}{2\tau}$.
In ferromagnets with strong polarization we are interested in, 
$\Delta\sim O(\epsilon_{F})$, and then neglecting $O(q/k_{F})$ 
contributions we may approximate 
$\epsilon^{*}_{\sigma,\pm}\sim \sigma 
\left(\Delta+\frac{\hbar^{2}k_{F\sigma}q}{2m}\cos\theta\right)
\pm  i\frac{\hbar}{2\tau}$,  where 
$\hbar k_{F\sigma}\equiv \sqrt{2m(\epsilon_{F}+\sigma\Delta)}$ is 
the Fermi momentum of the polarised electron.
The density of states estimated at the pole in the lower half plane, 
$\epsilon^{*}_{\sigma,-}$, has a opposite sign as the upper half 
plane, i.e., $N(\epsilon^{*}_{\sigma,\pm}) \sim \pm 
N(\sigma\Delta)\equiv \pm N_{\sigma}$. 
We therefore obtain
\begin{equation}
\int_{-\epsilon_{F}}^{\infty}d\epsilon N(\epsilon) F_{\theta}(\epsilon)
\simeq 2\pi\left(\frac{\tau}{\hbar}\right)^{3}\sum_{\sigma}N_{\sigma} 
\frac{1}{\left(\frac{\hbar^{2}k_{F\sigma}q}{2m}\cos\theta 
+\sigma\Delta\right)^{2}+\left(\frac{\hbar}{2\tau}\right)^{2} },
	\label{Jq3}
\end{equation}
where $N_{\sigma}\equiv(Vmk_{F\sigma}/2\pi^{2}\hbar^{2})$.
The integration over $\cos\theta$ in eq. (\ref{Jq1}) is then easily carried 
out to obtain
\begin{equation}
J_{q}=\frac{2m^{2}V\tau^{4}}{\pi\hbar^{8}} \sum_{\sigma} \frac{1}{q} 
\left[ 
\tan^{-1}\frac{2\tau}{\hbar}
       \left(\frac{\hbar^{2}k_{F\sigma}q}{2m}+\Delta\right)
+
\tan^{-1}\frac{2\tau}{\hbar}
       \left(\frac{\hbar^{2}k_{F\sigma}q}{2m}-\Delta\right)
\right].
	\label{Jq4}
\end{equation}
The conductivity (\ref{sigamcJ}) is then obtained by use of the expression of 
$A_{q}$,  eq. (\ref{aqdef}), 
($\sum_{q}|A_{q}|^{2}\cdots=(\pi/L)\int dq [\cosh^{2}(\pi q\lambda/2)]^{-1}\cdots$) 
as
\begin{equation}
  \sigma_{\rm c}=-\frac{e^{2}\Delta^{2}\tau^{2}}{8\pi\hbar^{3}L}\sum_{\sigma} 
  \int_{-\infty}^{\infty}\frac{dq}{q}\frac{1}{\cosh^{2}\frac{\pi}{2}q\lambda}
  \left[ 
\tan^{-1}\frac{2\tau}{\hbar}
       \left(\frac{\hbar^{2}k_{F\sigma}q}{2m}+\Delta\right)
+
\tan^{-1}\frac{2\tau}{\hbar}
       \left(\frac{\hbar^{2}k_{F\sigma}q}{2m}-\Delta\right)
\right].
	\label{sigmacJ1}
\end{equation}
Changing the variable $x\equiv\pi\lambda q/2$ we obtain (\ref{SIGMAC}).

To look into the asymptotic behaviors at large $\Delta$ it is useful to rewrite the result
by use of $\tan^{-1}x= \pm\pi/2-\tan^{-1}(1/x)$ for $x>(<)0$ as
\begin{equation}
\sigma_{\rm c}=-\frac{e^{2}}{4\pi\hbar}\tilDelta^{2} n_{\rm w}\sum_{\sigma} 
 \left[
 \pi \int_{\Lambda_\sigma}^{\infty}\frac{dx}{x}\frac{1}{\cosh^{2}x}
+
\int_{0}^{\infty}\frac{dx}{x}\frac{1}{\cosh^{2}x}
\tan^{-1} 
\frac{2 \till x}{(2\tilDelta)^{2}-(\till x)^{2}+1} \right],
	\label{sigmac2}
\end{equation}
where $\tilDelta\equiv \Delta\tau/\hbar$, 
$\till\equiv (2 l_{\sigma}/\pi\lambda)$, and $\Lambda_\sigma\equiv 2\tilDelta/\till 
=(\pi m \Delta\lambda/k_{F\sigma}\hbar^{2})$.
We consider the case $\tilDelta\gg1$ and $\Lambda_{\sigma}\gg1$, 
which would be satisfied for $d$ electrons in 3$d$ transition metals. 
In this case the inequality $(\tilDelta)^{2}\gg(\till)^{2},\till$
is satisfied if $\Delta/\epsilon_{F}$ is not too small.
Then the conductivity correction is approximated as
\begin{eqnarray}
\sigma_{\rm c}&=&-\frac{e^{2}}{4\pi\hbar}\tilDelta^{2} n_{\rm w}\sum_{\sigma} 
 \left[ 4\pi \int_{\Lambda_\sigma}^{\infty}\frac{dx}{x} e^{-2x}
+
\frac{\till}{2\tilDelta^{2}} \right]\int_{0}^{\infty}\frac{dx}{\cosh^{2}x}
 \nonumber\\
&=& -\frac{e^{2}}{4\pi^{2}\hbar} \frac{\Delta\tau^{2}}{m\lambda} n_{\rm w}
\sum_{\sigma} k_{F\sigma}
 \left[ 2\pi  e^{-2\Lambda_{\sigma}} +\frac{\hbar}{\Delta\tau} \right].
	\label{sigmac3}
\end{eqnarray}
In the case of a thick wall ($\lambda\gg k_{F}^{-1}$) with a finite 
$\tau$, the first term 
in (\ref{sigmac3}) is exponentially small and thus neglected.
The conductivity in this case is
\begin{equation}
	\sigma_{\rm  c} \simeq -\frac{e^{2}}{8\pi\hbar}n_{\rm w}\sum_{\sigma}
	\till\int_{0}^{\infty} \frac{dx}{\cosh^{2}x}
	= -\frac{e^{2}}{8\pi\hbar}n_{\rm w}\sum_{\sigma}
	\till   \;\;\;\;\;
	(\Delta\tau/\hbar, m\Delta\lambda/k_{F\sigma} \hbar^{2} \gg1).
	\label{sigmacferro}
\end{equation}

On the other hand if we take the limit of $\tau\rightarrow\infty$ 
first, the first term in (\ref{sigmac3}) becomes dominant and the 
result of Mori formula (eq. (\ref{cleanlimitrho})) is obtained. 
The result of Cabrera and Falicov\cite{Cabrera74}, obtained by
calculating the reflection coefficient in the absence of impurities, 
corresponds to this limit.

\section{Cooperons in ferromagnets}
\label{APPCooperon}

In this section we calculate the quantum correction to the conductivity 
in terms of the particle-particle ladder amplitude (Cooperon). 
In this section we discuss the Cooperon in ferromagnets with 
$\Delta\tau/\hbar \gg1$, 
taking into account the impurity scattering (eq. (\ref{Vimp})) but in 
the absence of domain wall.
Let us consider a correlation function  
\begin{eqnarray}
\lefteqn{	<< c_{\kv, n,\sigma}^{\dagger} c_{-\kv'+\qv, \np,\sigma'}
	    c_{-\kv+\qv, \np,\sigma'}^{\dagger} c_{\kv', n,\sigma} >>
=<c_{\kv, n,\sigma}^{\dagger}  c_{-\kv'+\qv, \np,\sigma'}
	    c_{-\kv+\qv, \np,\sigma'}^{\dagger}c_{\kv', n,\sigma} > }
\nonumber \\
&&  
+\frac{1}{2} \sum_{\kv_1,\kv_2,\qv'} v^2 
<\sum_{i,j}e^{i\qv' (\Xv_i-\Xv_j)}> _{\rm imp} \nonumber\\
&&\times
<c_{\kv, n,\sigma}^{\dagger}  c_{-\kv'+\qv, \np,\sigma'}
	    c_{-\kv+\qv, \np,\sigma'}^{\dagger} c_{\kv', n,\sigma}
	    c_{\kv_1, n,\sigma}^{\dagger} c_{\kv_2, n,\sigma} 
	    c_{-\kv_1+\qv', \np,\sigma'}^{\dagger} c_{-\kv_2+\qv', \np,\sigma'}> 
	    +\cdots
	   \nonumber\\ 
 &\equiv & -( G_{\kv, n,\sigma} G_{-\kv+\qv, \np,\sigma'} \delta_{\kv,\kv'}
	    +\Gamma_{nn'}^{\sigma\sigma'}(\kv,\kv',\qv)
	    G_{\kv, n,\sigma} G_{\kv', n,\sigma}
	    G_{-\kv'+\qv, n',\sigma'} G_{-\kv+\qv, \np,\sigma'} )
	    ,
	\label{Cooperondef}
\end{eqnarray}
where we have treated the impurity perturbatively and
double bracket denotes averaging over the electron states and 
impurities,  $<> _{\rm imp}$ being the average over impurity.
(In taking the impurity average, the self-energy processes are not 
included since they are already taken into account as a lifetime 
$\tau$ of the Green function.)
The function $\Gamma_{nn'}^{\sigma\sigma'}$ is defined by the last 
line and Cooperon denotes the singular part of 
$\Gamma_{nn'}^{\sigma\sigma'}$ for $q\sim0$, which is calcualted below.
By use of eq. (\ref{Cooperondef}) the current correlation function, 
eq. (\ref{Qdef}), 
which is related to the conductivity, is expressed as 
\begin{eqnarray}
	Q(i\omega_{\ell}) & = & -\left(\frac{e\hbar}{m}\right)^{2} 
	\frac{\hbar}{V\beta} \sum_{n\kv \sigma} \left[ 
     k_{z}^{2}G_{\kv n\sigma}G_{\kv,n+\ell,\sigma}  \right.
       \nonumber\\
&&	\left. +\sum_{\qv} k_{z}(-k+q)_{z} G_{\kv, n,\sigma} G_{\kv, n+\ell,\sigma}
	    G_{-\kv+\qv, n,\sigma} G_{-\kv+\qv,n+\ell,\sigma}  
	   \Gamma_{n,n+\ell}^{\sigma\sigma}(\kv'=-\kv+\qv,\qv)
	    \right]\nonumber\\
	  &\equiv &Q_0 +Q_{\rm 0q}.
	\label{fullQ}
\end{eqnarray}
The first term, $Q_0$ corresponds to the Boltzmann conductivity and the 
second term $Q_{\rm 0q}$ (described in Fig. \ref{FIGsigma0q}) 
turns out to describe the quantum effect. 
If the electron elastic mean free path is long (i.e., low impurity density) 
the second contribution $Q_{\rm 0q}$ 
is small but as $k_{F}l$ becomes smaller the 
this term becomes important (see in eq. (\ref{S0qS})).

We now show that $\Gamma_{n,n+\ell}^{\sigma\sigma}(\kv'=-\kv+\qv,\qv)$ 
contains a singular contribution at small $\qv$, which 
indicates the enhancement of the backward 
scattering due to the coherence, namely the weak localization of electron.
Cooperon denotes this singular contribution and is calcualted by summing 
over the ladder contribution shown in Fig. \ref{FIGCooperon}(a).
(There is another contribution to $\Gamma$ which is singular for 
$\qv\sim \kv+\kv'$, called a diffuson, represented by the 
diagram in Fig. \ref{FIGCooperon}(b), but we do not consider this 
process since it does not contribute to the conductivity because of 
the current vertex.)
The result of Cooperon is (we write   
$\Gamma_{n,n+\ell}^{\sigma\sigma}(\kv'=-\kv+\qv,\qv)$ simply as
$\Gamma_{n,n+\ell}^{\sigma\sigma}(\qv)$ below)
\begin{equation}
\Gamma_{nn'}^{\sigma\sigma'}(\qv)= n_{\rm i}v^{2} 
        \left[ 1+n_{\rm i}v^{2} I_{nn'}^{\sigma\sigma'}(\qv)+ 
	    (n_{\rm i}v^{2} I_{nn'}^{\sigma\sigma'}(\qv))^{2}+\cdots \right] 
	    \simeq \frac{ n_{\rm i}v^{2} }{1-n_{\rm i}v^{2} 
	    I_{nn'}^{\sigma\sigma'}(\qv)},
	    \label{Gammadef}
\end{equation}
where
\begin{equation}
	I_{nn'}^{\sigma\sigma'}(\qv)\equiv \sum_{\kv} 
	G_{\kv,n,\sigma} G_{-\kv+\qv,\np,\sigma'}.
	\label{Icooperondef}
\end{equation}
We consider the important case of small $q$ 
($q \lesssim l^{-1}$) , and then this function can 
be expanded with respect to $q$ as
\begin{equation}
  I_{nn'}^{\sigma\sigma'}(\qv)\simeq \sum_{\kv} 
	\left[ G_{\kv,n,\sigma} G_{\kv,n',\sigma'} 
	-\left(\frac{\hbar^{2}k_{z}q}{2m}\right)^{2} 
	\left\{  (G_{\kv,n,\sigma} G_{\kv,n',\sigma'} )^{2} 
	- G_{\kv,n,\sigma} (G_{\kv,n',\sigma'})^{3} 
	- (G_{\kv,n,\sigma} )^{3} G_{\kv,n',\sigma'} \right\}
	\right].	
	\label{Icooperonexpand}
\end{equation}
The summation over $\kv$ can be written in terms of the integration over the 
energy, $\epsilon$, from $-\epsilon_{F}$ to the infinity, and this integral 
is carried out in the same way as in eq. (\ref{Jq1}). 
The first term of (\ref{Icooperonexpand})is calculated as 
\begin{eqnarray}
	I_{+-}^{(1)\sigma\sigma'}
	  &\equiv & \sum_{\kv}  G_{\kv,n>0,\sigma} G_{\kv,n<0,\sigma'} \nonumber\\
	  &= & 
	\int_{-\epsilon_{F}}^{\infty} 
d\epsilon N(\epsilon) 
 \frac{1}
  {\left[ i\hbar\left(\frac{1}{2\tau}+\omega_{n} \right) 
                    -\epsilon+\sigma\Delta \right]}
 \frac{1}
  {\left[ -i\hbar\left(\frac{1}{2\tau}-\omega_{n'} \right) 
                    -\epsilon+\sigma'\Delta \right]} 
	         \nonumber\\
&\simeq& \frac{2\pi i}
{(\sigma-\sigma')\Delta
     +i\hbar\left(\frac{1}{\tau}+\omega_{n}-\omega_{n'}\right)} 
	      N(0), \label{Ikintegral}
\end{eqnarray}
where we have neglected higher order of 
$O(\epsilon_{F} \tau/\hbar)^{-1}$ and $O(\Delta/\epsilon_{F})$.
It is seen from this result that for the electrons with the same spin
($\sigma=\sigma'$), 
$I_{+-}^{(1)\sigma\sigma}
\simeq 2\pi N\tau/\hbar [1-(\omega_{n}-\omega_{n'})\tau]$
$+O((\omega_{n}-\omega_{n'})\tau)^{2}$ is large but for $\sigma=-\sigma'$, 
$I_{+-}^{(1)\sigma,-\sigma}$ is 
smaller by a factor of $(\Delta\tau/\hbar)^{-1}$.
In the case of $\sigma=\sigma'$ other terms in eq. 
(\ref{Icooperonexpand}) are calculated to obtain
\begin{equation}
I_{+-}^{\sigma\sigma}(\qv)\simeq  2\pi N\frac{\tau}{\hbar}
  [1-(Dq^{2}+\omega_{n}-\omega_{n'})\tau],
	\label{Iq2}
\end{equation}
where $D\equiv [(\hbar k_{F})^{2}\tau/3m^{2}]$.
In this case the sum of the ladder (\ref{Gammadef}) turns out to be 
singular at low energy (by use of $2\pi n_{\rm i}v^2 N(0)\tau/\hbar=1$);
\begin{equation}
	\Gamma_{+-}^{\sigma\sigma} \simeq
	\frac{n_{\rm i}v^{2}}{[Dq^{2}+\omega_{n}-\omega_{n'}]\tau }.
	\label{Gamma01}
\end{equation}
This singular behavior indicates the enhancement of the backward 
scattering amplitude due to the quantum 
interference\cite{Bergmann84,Lee85}.
On the other hand if $\sigma=-\sigma'$, 
$I^{\sigma,-\sigma}$ is small and thus $\Gamma_{+-}^{\sigma,-\sigma}$ 
is not important. 
For the case of both $n$ and $n'$ are positive or negative, 
it is easy to sea that the $\epsilon$-integral that corresponds to eq. 
(\ref{Ikintegral}) vanishes.
Therefore $\Gamma_{nn'}^{\sigma\sigma'}$ is important only if $nn'<0$ 
and $\sigma=\sigma'$. Other contributions are neglected below.

By use of (\ref{fullQ}) and (\ref{Gamma01}), we obtain the quantum 
correction to the conductivity as
\begin{eqnarray}
	\sigma_{0{\rm q}}&=&-\frac{1}{2\pi} \left(\frac{e\hbar}{m}\right) 
	\frac{1}{V} \frac{4\pi\hbar k_{F}^2}{3} N(0) 
	\left(\frac{\tau}{\hbar}\right)^3 \sum_{\qv,\sigma}\Gamma_{+-}^{\sigma\sigma}
	\nonumber\\
	&=& -\frac{2}{3\pi} \frac{e^2}{\hbar} \ell^2 \frac{1}{V} \sum_{\qv} 
	\frac{1}{Dq^2\tau}.
	\label{sigma0q}
\end{eqnarray}

In three-dimensions, $\frac{1}{V} \sum_{\qv} \frac{1}{Dq^2\tau} \simeq
\frac{3}{2\pi^2}\int_{L^{-1}}^{\ell^{-1}} \frac{dq}{\ell^2} =
\frac{3}{2\pi^2 \ell^3}$ and thus the ratio to the classical 
conductivity is obtained as
\begin{equation}
\frac{\sigma_{0{\rm q}}}{\sigma_0} =-\frac{3}{\pi}\frac{1}{(k_{F}\ell)^2} =
-\frac{3}{4\pi} \left(\frac{\hbar}{\epsilon_{F}\tau}\right)^2. 
\label{S0qS}
\end{equation}
One can see that $\sigma_{0{\rm q}}$ is a ``quantum'' correction, which 
vanishes in the limit of $\hbar\rightarrow0$.

\section{Summation over Matsubara frequency in eq. (\protect\ref{F2ADEF})}
\label{APPCFnsum}
Here we will carry out the summation over Matsubara frequencies in 
eq. (\protect\ref{F2ADEF}),
\begin{eqnarray}
J&\equiv& \frac{1}{\beta^{2}}\sum'_{n,n'} 
      (\tilGam(q,|n+l-n'|))^{2} (\tilGam(q,|n-(n'+\ell')|))^{2}
      \nonumber\\
&=& \frac{1}{\beta^{2}}\sum_{n,n'} 
\frac{1}
 {(Dq^{2}+|\omega_{n+\ell}-\omega_{n'}|)^{2}
 (Dq^{2}+|\omega_{n}-\omega_{n'+\ell'}|)^{2}}.
\end{eqnarray}
The summation here $\sum'_{n,n'}$ is restricted to the following three cases, 
where the Cooperons becomes important at small $q$;
\begin{equation}
 \begin{array}{ccc}
   {\rm I}  :  & n+\ell,n>0,         & n'+\ell',n'<0 \\	
   {\rm II} :  & n+\ell,n<0,         & n'+\ell',n'>0 \\	
   {\rm III}:  & n+\ell,n'+\ell'>0,  & n,n'<0 .
  \end{array}	
\end{equation}
Firstly the contribution from the case I is written as
\begin{eqnarray}
 J_{\rm I}&=&\frac{1}{\beta^{2}}\sum_{n=0}^{\infty}\sum_{n'=-\infty}^{-\ell'}
  \frac{1}
 {(Dq^{2}+\omega_{n+\ell}-\omega_{n'})^{2}(Dq^{2}+\omega_{n}-\omega_{n'+\ell'})^{2}}
   \nonumber\\
 &=&  
 \frac{1}{\beta^{2}}\sum_{n=0}^{\infty}\sum_{n''\equiv-(n'+\ell')=0}^{\infty}
 \frac{1}
 {(Dq^{2}+\omega_{n+\ell}+\omega_{n''+\ell'})^{2}(Dq^{2}+\omega_{n}+\omega_{n''})^{2}}.
\end{eqnarray}
We are interested in only the term proportional to 
$\omega_{\ell}\omega_{\ell'}$. 
Expanding $1/ (Dq^{2}+\omega_{n+\ell}+\omega_{n''+\ell'})^{2}$ with respect to 
$\omega_{\ell}$ and $\omega_{\ell'}$ we obtain this term as 
\begin{equation}
J_{\rm I}\simeq  \omega_{\ell}\omega_{\ell'}
\frac{1}{\beta^{2}}\sum_{n=0}^{\infty}\sum_{n''=0}^{\infty}
 \frac{1}
 {(Dq^{2}+\omega_{n}+\omega_{n''})^{6}}. 	
	\label{JI}
\end{equation}
The contribution from case II turns out to be the same; $J_{\rm II}=J_{\rm I}$.
Case III is treated as follows.
\begin{eqnarray}
 J_{\rm III}&=&\frac{1}{\beta^{2}}\sum_{n=-\ell}^{0}\sum_{n'=-\ell'}^{0}
  \frac{1}
 {(Dq^{2}+\omega_{n+\ell}-\omega_{n'})^{2}(Dq^{2}+\omega_{n'+\ell'}-\omega_{n})^{2}}
   \nonumber\\
 &=&  
 \frac{1}{\beta^{2}}
 \sum_{n''\equiv -n=0}^{\ell}\sum_{n'''\equiv n'+\ell'=0}^{\ell'}
 \frac{1}
 {(Dq^{2}-\omega_{n''-\ell}-\omega_{n'''-\ell'})^{2}
  (Dq^{2}+\omega_{n''}+\omega_{n'''})^{2}}
   .
\end{eqnarray}
By use of $\sum_{n=0}^{\ell}F(n)=\sum_{n=0}^{\infty}(F(n)-F(n+\ell))$ ($F(n)$ 
being any function) we obtain
\begin{eqnarray}
 J_{\rm III}&=&\frac{1}{\beta^{2}}
 \sum_{n\equiv =0}^{\infty}\sum_{n'=0}^{\infty}
 \left[ 
  \frac{1}
  {(Dq^{2}-(\omega_{n}+\omega_{n'})+\omega_{\ell+\ell'})^{2}
   (Dq^{2}+\omega_{n}+\omega_{n'})^{2}}
   \right. \nonumber\\
 &&  
 -\frac{1}
  {(Dq^{2}-(\omega_{n}+\omega_{n'})+\omega_{\ell'})^{2}
   (Dq^{2}+\omega_{n}+\omega_{n'}+\omega_{\ell})^{2}}
   \nonumber\\
 &&
 -\frac{1}
  {(Dq^{2}-(\omega_{n}+\omega_{n'})+\omega_{\ell})^{2}
   (Dq^{2}+\omega_{n}+\omega_{n'}+\omega_{\ell'})^{2}}
\nonumber\\
&&
\left.
 +\frac{1}
  {(Dq^{2}-(\omega_{n}+\omega_{n'}))^{2}
   (Dq^{2}+\omega_{n}+\omega_{n'}+\omega_{\ell+\ell'})^{2}}
  \right]
   .
\end{eqnarray}
Taking the term proportional to $\omega_{\ell}\omega_{\ell'}$ we obtain
$J_{\rm III}\simeq \omega_{\ell}\omega_{\ell'}J_{6}'$ where $J_{6}'$ is 
defined in eq. (\ref{Jmpdef}).
Adding these results we obtain eq. (\ref{F2aresult}).

\section{Calculation of diagrams with five and six Cooperons}
\label{APPCF56}
Here the processes containing five and six Cooperons 
(eqs. (\protect\ref{F5result})(\protect\ref{F6result})) are calculated.
Firstly let us evaluated the diagram Fig. \ref{FIGCF5}(a).
There are three different contributions shown in Fig. 
\ref{FIGCF5}(a-(i-iii)) with 
different configuration of $\delta\Gamma_{\rm w}$ and $\Gamma$.
The diagram Fig. \ref{FIGCF5}(a-i) is calculated as
\begin{eqnarray}
F_{5{\rm a(i)}} &=& \left(\frac{e\hbar}{m}\right)^{4} 
  \frac{1}{V^{2}}  
	\sum_{\kv,\kv',\kv'',\qv,\sigma} \frac{1}{\beta^{2}} \sum'_{n,n'}
	k_{z}(-k+q)_{z}k'_{z}(-k''+q)_{z}
	G_{\kv,n+\ell,\sigma} G_{\kv,n,\sigma} 
	G_{-\kv+\qv,n'+\ell',\sigma} G_{-\kv+\qv,n',\sigma} \nonumber\\
&&
\times  G_{\kv',n+\ell,-\sigma}G_{\kv',n,-\sigma}G_{-\kv'+\qv,n',-\sigma}
   G_{\kv'',n,-\sigma}G_{-\kv''+\qv,n',-\sigma}G_{-\kv''+\qv,n'+\ell',-\sigma}
   \nonumber\\
&& \times
   \delta\Gamma_{\rm w}(q,|n+\ell-n'|) \delta\Gamma_{\rm w}(q,|n-(n'+\ell')|)
         \Gamma(q,|n-n'|)
.
\end{eqnarray}
Here the summation $\sum'_{n,n'}$ is carried out in the two cases, 
\begin{eqnarray}
\begin{array}{ccc}
{\rm I:} & n+\ell,n>0, & n'+\ell', n'<0   \\
{\rm II:} &  n+\ell,n<0, & n'+\ell',n'>0 . 
\end{array} \label{IandII}
\end{eqnarray}
Summation over $\kv'$ and $\kv''$ gives rise to a contribution small 
at $q\rightarrow 0$; 
\begin{eqnarray}
\sum_{\kv'}k_{z}'
 G_{\kv',n+\ell,-\sigma}G_{\kv',n,-\sigma}G_{-\kv'+\qv,n',-\sigma}
&\simeq & 
\sum_{\kv'}k_{z}'
 G_{\kv',n+\ell,-\sigma}G_{\kv',n,-\sigma}G_{\kv',n',-\sigma}
  \left[ 1-\frac{\hbar^{2}(\kv'\cdot \qv)}{m}G_{\kv',n',-\sigma}\right]
 \nonumber\\
&&=-\frac{\hbar^{2}}{m}q_{z}\sum_{\kv'}k_{z}'
 G_{\kv',n+\ell,-\sigma}G_{\kv',n,-\sigma}(G_{\kv',n',-\sigma})^{2},
 \nonumber\\
&&\simeq  -\frac{q_{z}}{m\hbar} \frac{4\pi}{3}k_{F}^{2}N(0) \tau^{3},
\end{eqnarray}
for both cases I and II (in the last line we have neglected quantities of 
$O(\Delta/\epsilon_{F})^{2}$).
We thus obtain for case I 
\begin{eqnarray}
F_{5{\rm a(i)I}} &\simeq & -F_{0}\frac{2k_{F}^{2}q_{z}^{2}}{3m^{2}\tau^{3}}
w\left(\frac{r}{\lambda}\right)^{2}
   \nonumber\\
&& \times \sum_{n=0}^{\infty}\sum_{n'=-\infty}^{-\ell'}
        \frac{1}{(Dq^{2}+\omega_{n+\ell}-\omega_{n'})^{2}}
        \frac{1}{(Dq^{2}+\omega_{n}-\omega_{n'+\ell'})^{2}}
        \frac{1}{(Dq^{2}+\omega_{n}-\omega_{n'})}
.
\end{eqnarray}
By similar calculation as in Appendix \ref{APPCFnsum} and by use of 
$D\equiv \hbar^{2k_{F}^{2}}\tau/(3m^{2})$,  the term linear in both 
$\omega_{\ell}$ and $\omega_{\ell'}$ of $F_{5{\rm a(i)I}}$ results in
\begin{equation}
\frac{F_{5{\rm a(i)I}}}{\omega_{\ell}\omega_{\ell'}} 
 \simeq  -F_{0}[W\left({r}/{\lambda}\right)]^{2}
 \frac{2Dq_{z}^{2}}{\tau^{4}}8J_{7}.
	\label{F5aiI}
\end{equation}
Similarly, other contributions are calculated and the result of
$F_{5{\rm a}}\equiv 
\sum_{\alpha={\rm I,II}} F_{5{\rm a(i)\alpha}}
+\sum_{\alpha={\rm I,II}}F_{5{\rm a(ii)\alpha}}
+\sum_{\alpha={\rm I,II}}F_{5{\rm a(iii)\alpha}}$ 
is obtained as
\begin{eqnarray}
\frac{F_{5{\rm a}}}{\omega_{\ell}\omega_{\ell'}} 
 &\simeq&  -F_{0}[W\left({r}/{\lambda}\right)]^{2}
 \frac{2Dq_{z}^{2}}{\tau^{4}}J_{7}\left[ (8+8)+(4+10)+(10+4)\right]\nonumber\\
 &=&  -F_{0}[W\left({r}/{\lambda}\right)]^{2}
 \frac{2Dq_{z}^{2}}{\tau^{4}}44J_{7} .
	\label{F5a}
\end{eqnarray}
Diagram (a') is calculated as
\begin{eqnarray}
\frac{F_{5{\rm a'}}}{\omega_{\ell}\omega_{\ell'}} 
 &\simeq&  -F_{0}[W\left({r}/{\lambda}\right)]^{2}
 \frac{2Dq_{z}^{2}}{\tau^{4}}J_{7}\left[ 4+10+2+2+14+14\right] \nonumber\\
 &=&  -F_{0}[W\left({r}/{\lambda}\right)]^{2}
 \frac{2Dq_{z}^{2}}{\tau^{4}}46J_{7} .
	\label{F5a'}
\end{eqnarray}
Other diagrams (b)(c)(d) give rise to the same contributions as (a)+(a') and 
so the result of five Cooperons processes are obtained as
\begin{eqnarray}
\frac{F_{5}}{\omega_{\ell}\omega_{\ell'}} 
 &\simeq&  -F_{0}[W\left({r}/{\lambda}\right)]^{2}
 \frac{2Dq_{z}^{2}}{\tau^{4}}J_{7} (44+46)\times 4\nonumber\\
 &=&  -F_{0}[W\left({r}/{\lambda}\right)]^{2}
 720\frac{Dq_{z}^{2}}{\tau^{4}}J_{7} .
	\label{F5}
\end{eqnarray}

Diagrams of six Cooperons in Fig. \ref{FIGCF6} are calculated in the same 
way. 
For example, diagram (a) leads to 
\begin{eqnarray}
\frac{F_{6{\rm a}}}{\omega_{\ell}\omega_{\ell'}} 
 &\simeq&  -F_{0}[W\left({r}/{\lambda}\right)]^{2}
 \frac{(2Dq_{z}^{2})^{2}}{\tau^{4}}J_{8} 
      (14+7+10+11+14+7)\times 2\nonumber\\
 &=&  -F_{0}[W\left({r}/{\lambda}\right)]^{2}
 \frac{(2Dq_{z}^{2})^{2}}{\tau^{4}}126J_{8}  .
	\label{F6a}
\end{eqnarray}
Here bracket denotes the summation over six different ways of putting two 
$\delta \Gamma$'s and two $\Gamma$'s. 
Two factors of 2 in the first line are due to the two cases I and II.
There are other processes with different 
arrangement of $n'$ and $n'+\ell'$, corresponding to 
(a') in Fig. \ref{FIGCF4}, which give the equal contribution.Thus
\begin{equation}
\frac{F_{6{\rm aa'}}}{\omega_{\ell}\omega_{\ell'}} 
= -F_{0}[W\left({r}/{\lambda}\right)]^{2}
 \frac{(2Dq_{z}^{2})^{2}}{\tau^{4}}252J_{8}  .
	\label{F6aap}
\end{equation}
It turns out that (b),(c) leads to the same contribution and thus we obtain
\begin{equation}
\frac{F_{6}}{\omega_{\ell}\omega_{\ell'}} 
 \simeq  -F_{0}[W\left({r}/{\lambda}\right)]^{2}
 \frac{(Dq_{z}^{2})^{2}}{\tau^{4}}3024 J_{8} .
	\label{F6}
\end{equation}
\section{Calculation of $J_{m}$ and $J_{6}'$}
\label{APPJn}

The summation over the Matsubara frequency in the expression of 
$J_{m}$ and $J_{6}'$, eqs. (\ref{Jmdef}) and (\ref{Jmpdef}) is carried out 
here.
$J_{m}$ can be written by use of contour integration as
\begin{eqnarray}
J_{m} &=& \frac{1}{\beta^{2}} \sum_{n,n'=0}^{\infty} 
	\frac{1}{(D\qv^{2}+\omega_{n}+\omega_{n'})^{m}} \nonumber\\
&=& \left(\frac{-1}{2\pi i}\right)^{2} \int_{C_{+}}dz \int_{C_{+}}dz' 
 f(z)f(z') \frac{1}{[Dq^{2}-i(z+z')]^{m}} \nonumber\\
&=& -\left(\frac{1}{2\pi }\right)^{2} 
 \int_{-\infty}^{\infty} dz \int_{-\infty}^{\infty}dz' 
  f(z)f(z') \frac{1}{[Dq^{2}-i(z+z'+i0)]^{m}} 
,
\end{eqnarray}
where $C_{+}$ denotes the contour which surrounds the imaginary axis in the 
upper-half $z$-plane, and we have deformed the path to the path just above 
the real axis.
By use of partial integration we obtain
\begin{eqnarray}
J_{m} & = & \left(\frac{1}{2\pi }\right)^{2} \frac{1}{(m-1)(m-2)}
 \int_{-\infty}^{\infty} dz \int_{-\infty}^{\infty}dz' 
  \frac{df(z)}{dz}\frac{df(z')}{dz'} \frac{1}{[Dq^{2}-i(z+z'+i0)]^{m-2}} 
	\nonumber  \\
	 & = & \left(\frac{1}{2\pi }\right)^{2} \frac{1}{(m-1)(m-2)}
 \frac{1}{(Dq^{2})^{m-2}} .
	\label{Jmresult}
\end{eqnarray}
Similarly $J_{6}'$ is written as
\begin{eqnarray}
	J_{6}' &=& \left(\frac{1}{2\pi }\right)^{2}
\int_{-\infty}^{\infty} dz \int_{-\infty}^{\infty}dz' 
  f(z)f(z')\frac{d^{2}}{dzdz'}\left[ 
      \frac{1}{[Dq^{2}-i(z+z')]^{2}[Dq^{2}+i(z+z')]^{2}}  \right]
	\nonumber  \\
 & = & \left(\frac{1}{2\pi }\right)^{2} 
 \frac{1}{(Dq^{2})^{4}} .
\end{eqnarray}



%
%
\begin{figure}
\caption{A configuration of a domain wall in a mesoscopic 
wire.\label{FIGDW}}
\caption{The contributions to the Boltzmann conductivity 
which are the second order with 
respect to the interaction with the domain wall, denoted by wavy 
lines. Solid 
lines indicate the electron Green functions and the current vertex 
(expressed by crosses)
with wavy line represents $\delta J$.
\label{FIGdiagram}}
\caption{Shift of the electron density due to the second order 
interaction with the domain wall.
\label{FIGdeltan}}
\caption{ (a): Particle-particle ladder (Cooperon) in the absence of 
domain wall. 
Dotted line denotes the impurity scattering. 
This process is singular at $q\sim0$ if 
$\omega_n \cdot \omega_{n'} <0$.
(b): Particle-hole ladder (Diffuson) process. This process is also 
singular at $\qv \sim \kv+\qv'$ but does not contribute 
to the quantum correction to the conductivity.
\label{FIGCooperon}}
\caption{Quantum correction to the conductivity expressed in terms of 
Cooperon. Cooperon is defined by Fig. \protect\ref{FIGCooperon}(a), but
it is twisted here (maximally crossed diagram).
\label{FIGsigma0q}}
\caption{Effect of domain wall on the Cooperon to the second order. 
Hatched square denotes the particle-particle ladder (Cooperon).
(a): Self-energy type. This is the dominant process.
Process (b)  contains only one Cooperon and 
thus gives smaller contribution compared to (a).
Process (c) and the vertex correction type (d)(e)  include Cooperon(s) with 
different spins, and thus are suppressed in ferromagnets 
($\Delta\tau/\hbar\gg1$).
\label{FIGQC2nd}}
\caption{
Higher order orrection by the wall to the Cooperon
which represents a dephasing time by the wall. 
\label{FIGCooperonDW} }

\caption{Contributions to the conductance change due to the 
motion of the wall over a distance of $r$ at the lowest (fourth) 
order of domain wall interaction. Processes with four Cooperons are 
shown here. 
The domain wall is at $z=r$ and $z=0$ in outer and inner loop, respectively.
$n$ and $n+\ell$ denotes the Matsubara frequencies $\omega_{n}$ and 
$\omega_{n+\ell}$, respectively. (a) and (a') are different in the 
Matsubara frequencies of the internal loop. Current vertices are 
denoted by crosses.
\label{FIGCF4}}
\caption{Particle-particle ladder which includes the motion of the domain 
wall.
One of the electron line here interacts with the wall at $z=r$ and the other 
with the wall at $z=0$. 
\label{FIGdelG}}
\caption{Contributions to the conductance change containing five 
Cooperons. Although they contains more Cooperons their contributions are the 
same order as four Cooperons processes.
Diagrams (i)(ii)(iii) correspond to three different contributions which is 
obtained by cyclic replacement of $\delta \Gamma$'s and $\Gamma$.
In processes (b)-(d) only one of such contribution is shown.
\label{FIGCF5}}

\caption{Contributions to the conductance change containing six 
Cooperons. 
\label{FIGCF6}}
\end{figure}
\begin{figure}[hbt]
\caption{ The contour $C_{0}$ of integration in complex $z$-plane 
appeared in the 
summation over the Matsubara frequencies. Because of the function $f(z)$, 
there are poles on the imaginary axis at $z=(2n-1)\pi i/\beta$  where 
$n=0,\pm1,\pm2\cdots$.
$C_{1}$ is a deformed contour parallel to the real axis.
\label{FIGcontour}}

\caption{The contour of integration $C_{2}$ in complex $\epsilon$-plane. 
There is a cut on the real axis from $\epsilon=-\epsilon_{F}$ to $+\infty$,
because of the behavior of the density of states, 
$N(\epsilon)\propto(\epsilon+\epsilon_F)^{1/2} $ in 
three-dimensions. 
\label{FIGContour2}}

\end{figure}

\begin{thebibliography}{99}
\bibitem{RefKent} There has been, however, some exceptions; e.g., 
negative  $\Delta\rho_{\rm ani}$ has recently been observed in Fe 
wire\cite{Kent98}.
\end{thebibliography}
\end{document}